\nonstopmode

\documentclass[iop]{emulateapj}
\usepackage{color}
\usepackage{enumerate}
\usepackage{amsmath}
\usepackage[utf8]{inputenc}
\usepackage{graphicx}
\usepackage{natbib}
\usepackage{hyperref}
\newcommand{\curlyc}{{\cal C}}

\begin{document}

\title{The Evolution of Post-Starburst Galaxies from $z\sim1$ to the Present}
\shorttitle{Evolution of Post-Starburst Galaxies}
\shortauthors{Pattarakijwanich et al.}

\author{Petchara Pattarakijwanich\altaffilmark{1,*,$\dagger$}, Michael A. Strauss\altaffilmark{1}, Shirley Ho\altaffilmark{2}, Nicholas P. Ross\altaffilmark{3}}
\altaffiltext{1}{Department of Astrophysical Sciences, Princeton University, Peyton Hall, 4 Ivy Lane, Princeton, NJ 08544, USA}
\altaffiltext{2}{McWilliams Center for Cosmology, Department of Physics, Carnegie Mellon University, 5000 Forbes Avenue, Pittsburgh, PA 15212}
\altaffiltext{3}{Department of Physics, Drexel University, 3141 Chestnut Street, Philadelphia, PA 19104, USA}
\altaffiltext{*}{Currently a KIAA Postdoctoral Fellow at the Kavli Institute for Astronomy and Astrophysics, Peking University, 5 Yi He Yuan Road, Hai Dian District, Beijing, 100871, China}
\altaffiltext{$\dagger$}{ppattara@pku.edu.cn}

\begin{abstract}

Post-starburst galaxies are in the transitional stage between blue,
star-forming galaxies and red, quiescent galaxies, and therefore hold important
clues for our understanding of galaxy evolution. In this paper, we
systematically searched for and identified a large sample of post-starburst
galaxies from the spectroscopic dataset of the Sloan Digital Sky Survey (SDSS)
Data Release 9. In total, we found more than 6000 objects with redshifts
between $z\sim0.05$ and $z\sim1.3$, making this the largest sample of
post-starburst galaxies in the literature. We calculated the luminosity
function of the post-starburst galaxies using two uniformly selected
subsamples: the SDSS Main Galaxy Sample and the Baryon Oscillation
Spectroscopic Survey CMASS Sample. The luminosity functions are reasonably fit
by half-Gaussian functions. The peak magnitudes shift as a function of redshift
from $M\sim-23.5$ at $z\sim0.8$ to $M\sim-20.3$ at $z\sim0.1$. This is
consistent with the downsizing trend, whereby more massive galaxies form
earlier than low-mass galaxies. We compared the mass of the post-starburst
stellar population found in our sample to the decline of the global
star-formation rate and found that only a small amount ($\sim1\%$) of all
star-formation quenching in the redshift range $z=0.2-0.7$ results in
post-starburst galaxies in the luminosity range our sample is sensitive to.
Therefore, luminous post-starburst galaxies are not the place where most of the
decline in star-formation rate of the universe is happening.

\end{abstract}

\maketitle

\section{Introduction}
\label{sec:introduction}

Post-starburst galaxies were first identified by \citet{Dressler_Gunn_1983} as
a distinct class of galaxies, whose main spectroscopic characteristics are
strong Balmer absorption lines with a lack of emission lines due to star
formation. The physical explanation of these properties is that these galaxies
had a large burst of star formation about $10^{8-9}$ years before the time of
observation that suddenly was quenched, with very little ongoing star
formation. This stellar population is old enough that the short-lived O- and
B-stars have all died (thus no nebular emission lines), and young enough that
the optical spectrum is dominated by A-stars (thus the strong Balmer
absorption). Given the stellar population properties, one possible explanation
is that these objects are in a transitional stage between star-forming blue
cloud galaxies and quiescent red sequence galaxies.

The global star formation rate over the entire universe is found to have
dropped by about an order of magnitude since redshift $z\sim1$
\citep{Madau_Dickinson_2014}. This implies that the galaxy population changed
drastically between redshift $z\sim1$ and the present time. The heavily
star-forming galaxies that are abundant at high redshift must be quenched and
evolve to the red, passive elliptical galaxies. If this quenching is sudden,
then they will pass through the post-starburst phase in between. Therefore, we
expect that the post-starburst galaxy population in this redshift range would
reflect this change in the global star formation rate. In order to understand
this, one needs a large, statistical sample of post-starburst galaxies that are
selected with a uniform and well-understood selection method across this large
range of redshift.

The terminology in post-starburst galaxy studies is rather rich and confusing.
Two common names are ``K+A'' and ``E+A'', referring to the fact that the
spectra of this kind of galaxy can be decomposed into spectra of ``K''- and
``A''-stars (and ``E'' stands for elliptical galaxy, whose spectrum is
dominated by K-giants). Some authors have also defined a number of subclasses
such as E+a \citep{Choi_etal_2009} and post-quenching
\citep{Quintero_etal_2004}. They are also closely related to the ``green
valley'' galaxies, lying intermediate between the red sequence and blue cloud,
although not all green valley galaxies have the spectroscopic signature of
post-starburst galaxies. In this paper, we will use a specific working
definition of post-starburst galaxies based on quantities measured from optical
spectra, but otherwise we will make no particular distinction between these
names and just call such objects collectively as ``post-starburst''.

Two main scenarios have been proposed to explain the formation of
post-starburst galaxies, which predict different detailed properties of these
galaxies. While it is likely that both mechanisms contribute to the formation
of post-starburst galaxies to some degree, the relative importance of the two
mechanisms is far from clear. Moreover, the relative importance of each
mechanism can potentially be a function of redshift and galaxy mass.

The first scenario is a cluster-related mechanism such as ram-pressure
stripping. In this picture, gas-rich star-forming galaxies fall into galaxy
clusters and their gas is removed by the interaction with the hot intra-cluster
medium, suddenly quenching their star formation \citep{Gunn_Gott_1972,
Balogh_etal_2000}. According to this scenario, the resulting post-starburst
galaxies would be found predominantly in dense environments such as galaxy
groups and clusters. They would still resemble disk galaxies, since nothing
apart from gas loss would disturb their morphologies. This idea was first
proposed by \citet{Dressler_Gunn_1983} because the first post-starburst
galaxies they found were in clusters, and they therefore hypothesized that
these objects lay exclusively in overdense regions. It was only found later
that these objects are also present in the field \citep{Zabludoff_etal_1996,
Quintero_etal_2004, Blake_etal_2004}. Interaction with the intra-cluster medium
is also proposed as a possible way to form S0 galaxies.

An alternative scenario is that post-starburst galaxies are associated with
galaxy interactions, mergers or AGN feedback. In this scenario, when gas-rich
galaxies go through a merger phase, the gas is disturbed and collapses to form
stars, leading to a large scale starburst. The same disturbance also funnels
the gas to the galactic center, resulting in both a central starburst and AGN
activity. The galaxy then either runs out of gas and stops forming stars, or
the rest of the gas receives enough heating from either supernova or AGN
feedback that it becomes too hot to collapse further or is expelled altogether.
Either way, the galaxy goes through a starburst phase that stops quickly. This
picture makes a number of predictions about the properties of post-starburst
galaxies. The first is that it naturally explains the observations of
post-starburst galaxies in lower density regions such as poor groups and the
field. Their morphologies should show prominent signs of recent interactions
such as tidal tails \citep{Zabludoff_etal_1996, Yang_etal_2004, Yang_etal_2008,
Blake_etal_2004, Tran_etal_2004, Goto_2005}. Post-starburst galaxies are also
expected to generally coincide with AGN, since both starburst and AGN activity
are likely triggered by the same mechanisms. Also, the post-starburst stellar
population is predicted to be centrally concentrated in their galaxies because
the gas is driven toward the center in a merger, making the post-starburst
galaxies appear more compact than elliptical galaxies at similar redshift.
Observations show that this indeed is the case \citep{Pracy_etal_2012,
Swinbank_etal_2011, Whitaker_etal_2012}.

The picture that gas-rich mergers drive both star formation and AGN activity is
supported by theoretical work as early as, for example,
\citet{Sanders_etal_1988} and \citet{Mihos_Hernquist_1994}. More recent
simulation papers that study this scenario include, for example,
\citet{Hopkins_etal_2006, Hopkins_etal_2008, Bekki_etal_2001, Bekki_etal_2005,
Bekki_etal_2010} and \citet{Snyder_etal_2011}. Their models suggest that the
spectrum of a merging system starts with strong Balmer absorption with a
moderate amount of emission lines, then goes through a post-starburst phase
when the emission lines fade out. After about 1 Gyr, the galaxy becomes
dominated by an old stellar population.

There are also other models that attempt to explain this post-starburst
phenomenon. \citet{Poggianti_Wu_2000} and \citet{Miller_Owen_2001} suggest that
post-starburst galaxies are in fact dust-enshrouded starburst galaxies. In this
picture, the ongoing star formation is hidden behind a large amount of dust and
gas and therefore the emission lines are completely extincted. Balmer
absorption lines, on the other hand, originate from A-stars, which are
long-lived enough to migrate out of the star-forming gas cloud and become
visible. This idea can be tested at far-infrared or radio continuum
wavelengths, at which one can measure the star formation rate independent of
dust and gas extinction. Various authors have carried out this test and the
conclusion is that even though a small number of post-starburst galaxies may
include dust-enshrouded starbursts, this model does not explain the majority of
the post-starburst galaxy population \citep{Miller_Owen_2001, Goto_2004,
Chang_etal_2001, Nielsen_etal_2012}.

In order to study post-starburst galaxies and understand the relative
importance of different star-formation quenching mechanisms, the first
important task is the selection of the sample itself. Post-starburst galaxies
are typically selected spectroscopically, with the requirement that the Balmer
absorption be strong. The H$\delta$ line is commonly used for this purpose
because emission from star formation tends to be quite weak in this line,
unlike H$\alpha$ or H$\beta$. It also does not have other major lines in its
immediate vicinity, making the equivalent width measurement easy. The nebular
emission lines, commonly represented by H$\alpha$ and [OII]3727, are required
to be weak. For objects at higher redshift, H$\alpha$ drops out of the
wavelength coverage of optical instruments and therefore the selection is
usually done on the H$\delta$ and [OII] lines alone. There are also a number of
alternative methods such as spectral template fitting
\citep{Quintero_etal_2004}, Principal Component Analysis \citep{Wild_etal_2007,
Wild_etal_2014} and using UV-optical colors \citep{Choi_etal_2009}.

A large sample is important to statistically represent the whole population.
Large spectroscopic surveys have enabled the selection of samples of
post-starburst galaxies containing hundreds of objects at low redshift
\citep{Zabludoff_etal_1996, Goto_2005, Goto_2007, Wild_etal_2007,
Quintero_etal_2004}, and a few tens at higher redshift \citep{Blake_etal_2004,
Yan_etal_2009, Vergani_etal_2010}, compared to a handful of objects in earlier
studies. With these samples, various global properties of the samples have been
studied, such as the luminosity function \citep{Blake_etal_2004}, their
clustering \citep{Krause_etal_2013} and quantification of their environments
\citep{Zabludoff_etal_1996, Galaz_2000, Blake_etal_2004, Balogh_etal_2005,
Hogg_etal_2006, Yan_etal_2009, Poggianti_etal_2009}.

More detailed information can also be obtained for a small number of
post-starburst galaxies through follow-up observations. Detailed morphological
studies using HST have been carried out by \citet{Belloni_1997,
Caldwell_etal_1999, Tran_etal_2003, Tran_etal_2004, Blake_etal_2004,
Yang_etal_2004, Yang_etal_2008} and \cite{Cales_etal_2011}, who have shown that
they show a range of morphologies. Spatially resolved spectroscopic data,
either through long slit or integral field units, have been obtained and used
to study the dynamical properties \citep{Caldwell_etal_1996} and the spatial
distribution of the post-starburst stellar population in individual galaxies
\citep{Norton_etal_2001, Goto_2005, Yagi_Goto_2006, Pracy_etal_2005,
Pracy_etal_2009, Pracy_etal_2010, Pracy_etal_2012, Goto_etal_2008,
Swinbank_etal_2011}.

In this paper, we tackle this rich and complicated problem at the key
limitation of previous studies - small sample size - by first identifying a
large number of post-starburst galaxies from the Sloan Digital Sky Survey
(SDSS) database over a very broad range of redshift. In Section \ref{sec:data}
we introduce the SDSS data, and give some technical background. In Section
\ref{sec:selection} we describe the method used to identify the post-starburst
galaxies from this large spectroscopic dataset. The properties of the resulting
sample is then described in Section \ref{sec:properties}. The luminosity
function analysis of this sample, and its relation to the global star-formation
rate, are then explained in Section \ref{sec:luminosity_function}. The
discussion and conclusion of the results are then presented in Sections
\ref{sec:discussion} and \ref{sec:conclusion}.

\section{Data}
\label{sec:data}

\subsection{SDSS and BOSS Surveys}

The Sloan Digital Sky Survey \citep{York_etal_2000} has taken both imaging and
spectroscopic data over more than a quarter of the sky, using a dedicated 2.5m
telescope situated at the Apache Point Observatory in New Mexico
\citep{Gunn_etal_2006}. The imaging data are taken in the five-band filter
system - $ugriz$ - defined in \citet{Fukugita_etal_1996} with the mosaic CCD
camera described in \citet{Gunn_etal_1998}. The raw imaging data are processed,
calibrated and cataloged by algorithms described in \citet{Lupton_etal_2002}. A
subset of detected sources are selected for spectroscopic follow-up by a number
of selection criteria based on observed magnitudes and colors. These targets
are then assigned to spectroscopic tiles \citep{Blanton_etal_2003} and their
spectra obtained with the spectrograph described in \citet{Smee_etal_2013}. In
the first two phases of the survey (SDSS-I, SDSS-II) the selection was
concentrated into three categories: a magnitude-limited sample of galaxies; the
Main Galaxy Sample \citep{Strauss_etal_2002}, luminous red galaxies (LRG) at
higher redshift \citep{Eisenstein_etal_2001} and quasars
\citep{Richards_etal_2002}. Spectra obtained in these phases cover the
wavelength range between $3800-9200$\AA\, at resolution of $R\sim2000$, with
typical signal-to-noise ratio per resolution element $>4$ for objects brighter
than $g=20.2$. The spectra are both wavelength- and
spectrophotometry-calibrated. In the third phase of the survey, SDSS-III
\citep{Eisenstein_etal_2011}, the BOSS program \citep{Dawson_etal_2013} was
carried out with an upgraded spectrograph (wavelength coverage $3600-10400$\AA,
with signal-to-noise ratio $>5$ at $i=21$). This program concentrated on
galaxies and quasars at higher redshift for measurement of Baryon Acoustic
Oscillations. The spectra observed are classified with redshifts measured by
the spectral classification algorithm explained in \citet{Bolton_etal_2012}. In
addition, for the first two phases of the SDSS survey, an alternative spectral
classification was performed \citep{SubbaRao_etal_2002}. In this work we based
our study on all available spectra in SDSS-I/II and BOSS spectra in SDSS-III up
to the Data Release 9 of the survey (SDSS DR9; \citealt{Ahn_etal_2012}).

\subsection{SDSS Magnitude System}
\label{sec:sdss_magnitude}

For each SDSS object, the inverse hyperbolic sine magnitude (asinh;
\citealt{Lupton_etal_1999}) in the AB system in all five bands is measured from
raw imaging data by a number of different methods, resulting in a variety of
magnitudes that we will find ourselves using throughout this paper. Fiber
magnitude measures the flux contained within the spectrograph fiber. Petrosian
magnitude is designed to measure galaxy photometry optimally
(\citealt{Petrosian_1976, Blanton_etal_2001, Yasuda_etal_2001} and
\citealt{Strauss_etal_2002}). The PSF magnitude measures the flux of a point
source, using the PSF as a matched filter. Model and cModel magnitudes measure
the flux of a galaxy by fitting surface brightness profiles to the image. The
Spectro magnitude measures the brightness from the spectra, relying on the
excellent spectrophotometric calibration of SDSS
\citep{Adelman-McCarthy_etal_2008}. In later analysis, we distinguish and
convert between these magnitude systems. More details on this can be found in
the SDSS website \url{http://www.sdss.org/dr12/algorithms/magnitudes/}.

\subsection{Small Sample of Post-Starburst Galaxies}
\label{sec:known_sample}

We assembled a small sample of known post-starburst galaxies at both low and
high redshift, which will be used to test and tune the selection criteria that
we will introduce in subsequent sections. This sample contains the 564
low-redshift ($z_\mathrm{median} \sim 0.1$) post-starburst galaxies found in
the 5th Data Release of the Sloan Digital Sky Survey
\citep{Adelman-McCarthy_etal_2007} by \citet{Goto_2007}. This sample was
selected from the main galaxy catalogue at redshift $z<0.33$ such that the
H$\alpha$ line is well within the wavelength coverage of the spectra. It was
selected by the equivalent width cuts H$\delta$ EW $>$ $5.0$\AA, H$\alpha$ EW
$>$ $-3.0$\AA \, and OII$\lambda$3727 EW $>$ $-2.5$\AA , where positive value
of equivalent width corresponds to absorption and negative value to emission.
We add to this about 100 intermediate- and high-redshift objects (redshift $z
\sim 0.4 - 0.8$) that were mostly found serendipitously from visual inspection
of SDSS and BOSS spectra.

\section{Selection of post-starburst galaxies}
\label{sec:selection}

\subsection{Spectral Templates}

\begin{figure}[h]
\begin{center}
\includegraphics[trim=30 10 10 25,clip,width=0.475\textwidth]{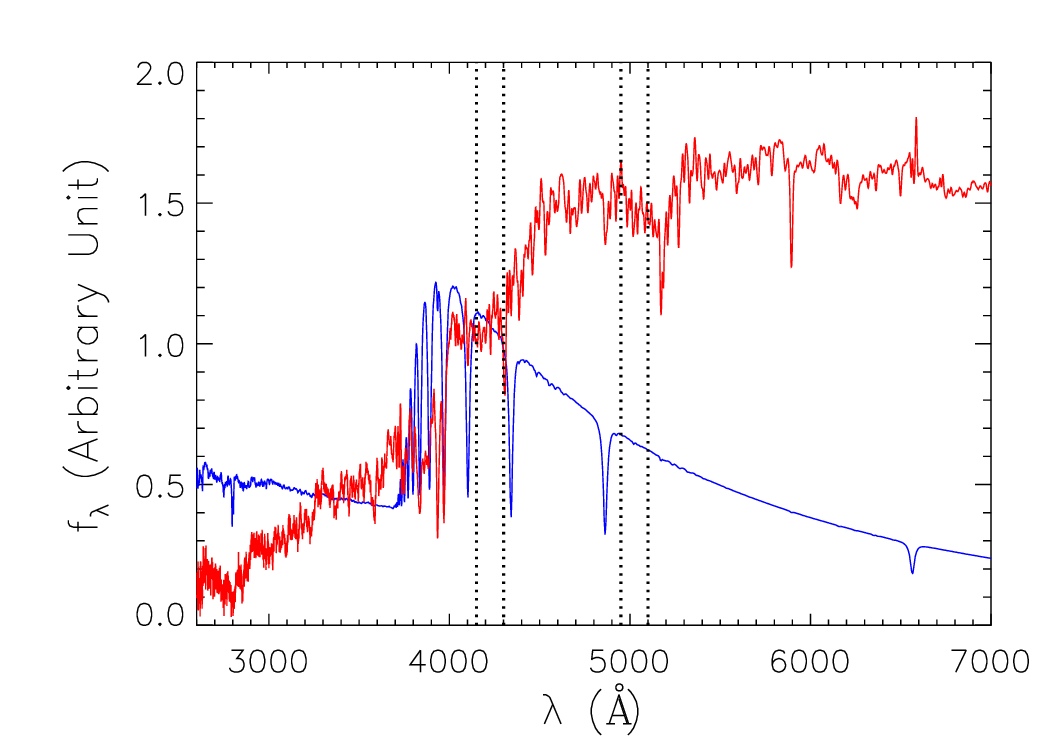}

\caption{This plot shows the spectra used as fitting templates for ``A'' (blue)
and ``K'' (red) components of the galaxy population. The ``A''-template is the
spectrum of Vega, which is representative of generic A0V stars, from
\citet{Bohlin_Gilliland_2004}. The ``K''-template is the composite spectrum of
Luminous Red Galaxies (LRGs) in SDSS \citep{Eisenstein_etal_2003}. The fits are
carried over the wavelength range 3600-4400\AA\, and the templates are
normalized such that the integrated fluxes over this range are the same. The
vertical dotted lines show the wavelength ranges 4150-4300\AA\, and
4950-5100\AA\, (hereafter called [4200] and [5000] respectively) corresponding
to our fiducial bands used in later sections.}

\label{fig:template}
\end{center}
\end{figure}

The selection method we use is based on template fitting, following
\citet{Quintero_etal_2004}. In this section we describe the templates we use.
As suggested by the name ``E+A'' or ``K+A'', these objects' spectra can be
decomposed into a linear combination of the spectrum of an old stellar
population (the ``E'' or ``K'' component), and the spectrum of an A-type star.
We use the old stellar population template of \citet{Eisenstein_etal_2003}.
This is the composite spectrum of Luminous Red Galaxies in SDSS, which is
dominated by the light of K-type giants. The spectrum of Vega (an A0V star)
from \citet{Bohlin_Gilliland_2004} is used to represent the ``A''-template.
Both ``K'' and ``A'' templates are normalized such that the total fluxes over
the wavelength range 3600-4400 \AA \, are the same, as shown in Figure
\ref{fig:template}. We emphasize that this template fitting approach is only a
crude fit to the underlying stellar population. A proper and more detailed
analysis of the stellar populations in our sample will be carried out in future
work.

\subsection{Selection Methods}

\subsubsection{Redshift Range Determination}

The SDSS main spectral classification pipeline provides redshifts and
classifications for all spectra based on the Principal Component Analysis (PCA)
technique \citep{Bolton_etal_2012}. This pipeline redshift is reliable for
object types that are common and well represented in the PCA templates. This is
not always the case for the post-starburst galaxies that we try to select,
especially at high redshift. In the preliminary stage of this work we found
that a number of known post-starburst galaxies, especially at high redshift,
have wrong redshifts and are misclassified. Since we desire that the final
sample be as complete as possible, we do not assume that the redshift value
given by the pipeline is reliable, and will determine the redshift from the
spectrum as if the redshift is not known a priori. For this reason we allow
redshift to be a free parameter when we perform the template fitting. The
redshift is constrained such that the rest-frame wavelength range of the
template, 3600-4400\AA, is contained within the observed spectra, corresponding
roughly $z=0.05-1.1$ for the SDSS wavelength coverage (3800-9200\AA) and
$z=0-1.35$ for the BOSS wavelength coverage (3600-10200\AA).

However, in SDSS-I/II, an alternative, independent spectral classification
pipeline is available \citep{SubbaRao_etal_2002}. This pipeline disagrees in
redshift with the main pipeline significantly for about 2\% of the objects.
Even though both of these pipelines can be individually incorrect, when the two
agree the redshift is highly likely to be correct. So when this happens (more
specifically, when the two redshifts differ by less than 0.005) we do not
perform the template fitting over the full redshift range, but only around a
narrow range $\Delta z = 0.008$ around the average of the two to find the local
optimum. For BOSS, since this alternative pipeline is not available, the
redshifts for BOSS galaxies are never assumed to be correct a priori (see the
discussion in Section \ref{sec:compare_pipeline_redshift}).

\subsubsection{Redshift Determination}
\label{sec:redshift_determination}

Once the redshift range to be fit over is determined, the template fitting is
performed at every value of redshift in the range with uniform spacing $\Delta
z = 0.001$. For each redshift value two models are fitted to the wavelength
corresponding to 3600-4400 \AA \, rest-frame wavelength, shifted to this value
of redshift. The first model is to fit a linear combination of the
``K''-template, ``A''-template and a second-order polynomial in wavelength as
continuum. This is written mathematically as

\begin{equation}
y_{\mathrm{Model1}} = C_{K}K(\lambda) + C_{A}A(\lambda) + 
                      C_{0} + C_{1}\lambda + C_{2}\lambda^2
\label{eq:model_fit}
\end{equation}

\noindent where $K(\lambda)$ and $A(\lambda)$ correspond to the ``K'' and ``A''
templates. The continuum is included to take into account other possible
slowly-varying effects such as AGN continuum, dust extinction,
spectrophotometric errors and template errors. The second model is to fit only
the second-order polynomial in wavelength, without the ``K'' and ``A''
templates.

For each fit, the value of the $C$ parameters are determined by the standard
minimum $\chi^2$ fitting method, with each pixel weighted by the inverse square
of the per-pixel error output by the spectroscopic pipeline. This method can
yield negative, unphysical values of $C_K$ or $C_A$. There are other classes of
fitting method that can constrain parameters to be positive-definite. However,
considering the ease of implementation and the fast computational speed
(especially since we need to run this process on the whole SDSS and BOSS
spectroscopic database, consisting of more than 2 million spectra), we use this
simple $\chi^2$ fit and tolerate this issue. This is not a problem since visual
inspection (see below) is done at the end to ensure the quality of all fits
anyway. Generally, it is also found that the polynomial component is usually
small compared to the ``K'' and ``A'' components, thus the spectra are
typically well represented by a linear combination of the two spectral
templates.

For each redshift, we use the difference in $\chi^2$ values of the two models,
$\Delta\chi^2 = \chi^2_{\mathrm{Model2}} - \chi^2_{\mathrm{Model1}}$, as an
indicator of a good template fit. $\Delta\chi^2$ indicates how much better the
fit is for the model with templates included, compared to the pure polynomial
model. A large value of $\Delta\chi^2$ could only happen if at least one of the
``K'' or ``A'' template is a good representation of the spectrum at that
redshift. We take the redshift with the maximum value of $\Delta\chi^2$ as the
true redshift of the object.

For true post-starburst galaxies we found that the value of $\Delta\chi^2$ is
at least around 40-50 (much larger than $\Delta\chi^2\sim2$ expected by
increasing the number of parameters by 2), and up to a few thousand for objects
with strong post-starburst features with good signal-to-noise ratio. Values of
$\Delta\chi^2$ lower than this are usually associated with low signal-to-noise
spectra and it is usually unclear whether the post-starburst spectral features
are actually detected, therefore they are removed in the visual inspection
process.

\subsubsection{Identification of Post-starburst objects}

To quantify the contribution of the ``A''-spectrum component to the fit in
order to identify post-starburst galaxies, we introduce a parameter,
$A/\mathrm{Total}$, defined as $A/(K+A+P)$ where $K$, $A$ and $P$ are the
integrated fluxes in the wavelength range 4150-4300\AA\, (hereafter called
[4200]) of the ``K'', ``A'' and the polynomial components of the model fit
described above. This wavelength range corresponds to the continuum between the
two Balmer absorption lines H$\gamma$ and H$\delta$ (Figure \ref{fig:template})
which are prominent features in the A-star spectrum. We require that this
$A/\mathrm{Total}$ ratio be greater than 0.25 for an object to be considered a
candidate post-starburst galaxy. This particular cut at 0.25 is determined such
that the selection algorithm recovers objects in the sample of known
post-starburst galaxies (Section \ref{sec:known_sample}) with reasonably strong
spectral features.

\subsubsection{Equivalent Width Cut; Visual Inspection}
\label{sec:ew_cut}

We also put requirements, adopted from \citet{Goto_etal_2003}, on the
equivalent widths of two key spectral lines. We require that the rest-frame
equivalent widths satisfy H$\delta$EW $>$ 4.0\AA \, and [OII]$\lambda$3727EW
$>$ -2.5\AA, where a positive value in equivalent width corresponds to
absorption. The requirement on strong Balmer absorption guarantees a
substantial contribution from the A-star component, while the lack of [OII]
emission indicates an insignificant amount of ongoing star formation. The
requirement on the H$\delta$ equivalent width is somewhat redundant with the
requirement on A/Total ratio, since they both measure the contribution of the
A-star component. We still include both to ensure that the high A/Total ratio
is not only due to the good fit to the continuum by the A-star template. These
cuts are less conservative than the cuts used in \citet{Goto_2007} (H$\delta$EW
$>$ 5.0\AA). This choice is made so that the sample is as inclusive as possible
before all candidates are visually inspected.

The equivalent widths are calculated by summing the flux density over the line.
A linear function is fit to the red and blue sidebands of the line to define
the continuum region, then the excess over this continuum in the line region is
used to calculate the equivalent width. The values of rest-frame wavelengths
used to define the red and blue sidebands and the bandpass of relevant lines
(including H$\alpha$, which will be used in following sections) are shown in
Table \ref{tab:ew_wavelength}. We did not apply any cut on the signal-to-noise
ratio of the equivalent width measurements. We will quantify the effects of
limited signal-to-noise ratio when we derive the luminosity function in Section
\ref{sec:luminosity_function}.

It should be noted that this selection method is designed to select only those
post-starburst galaxies in which the starburst had occurred at least several
tens Myr previously. Significantly younger post-starburst galaxies will have
some [OII]$\lambda$3727 emission or weaker H$\delta$ absorption, and thus can
be excluded from the sample. We will discuss the effects of this incompleteness
below.

\begin{figure*}
\begin{center}
\includegraphics[trim=20 15 25 10,clip,width=\textwidth]{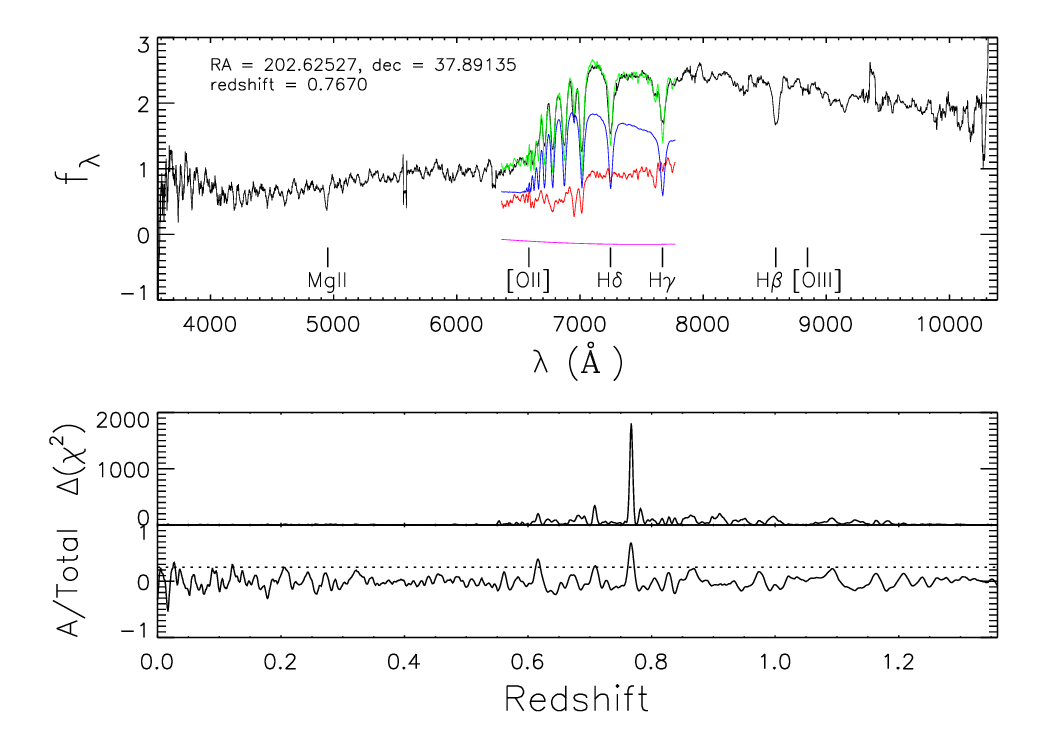}

\caption{The template fitting procedure is demonstrated in this plot. The
object shown is a post-starburst galaxy from BOSS at
$\mathrm{RA}=202.62527^\circ$ and $\mathrm{DEC}=37.89135^\circ$ with redshift
$z=0.7670$, observed on plate 3984, fiber 233 and MJD 55333. For this object,
the pipeline redshift \citep{Bolton_etal_2012} is accurate. The top panel shows
the spectrum of this object, smoothed for increased visibility. Overplotted are
the best-fit model (green) with its ``K'', ``A'' and polynomial components
(red, blue and magenta respectively) at the correct redshift. Positions of
several major lines are shown. For this object, there is no emission in
[OII]$\lambda$3727 as required by our selection criteria, while the H$\alpha$
line is outside the spectral coverage. The bottom panels show the parameters
$\Delta(\chi^2)$ and A/Total of the fit done at all values of assumed redshift
within the allowed range. The candidate redshift is where the inclusion of
templates in the fit maximally improves the fit, resulting in the peak in
$\Delta(\chi^2)$. At this redshift the A/Total ratio is required to be higher
than the threshold value of 0.25, represented by the horizontal dotted line, in
order to enter the sample. After passing this criterion the fit is visually
inspected before the object is included in the final sample.}

\label{fig:fitprocedure}
\end{center}
\end{figure*}

\begin{table*}
\begin{center}
\begin{tabular}{|c|c|c|c|c|}
\hline
Line & Bandpass & Blue Sideband & Red Sideband & Reference \\
\hline
[OII]$\lambda$3727 & 3716.30-3738.30\AA &
  3696.30-3716.30\AA & 3738.30-3758.30\AA & \citet{Yan_etal_2006} \\
H$\delta$ & 4083.50-4122.25\AA &
  4041.60-4079.75\AA & 4128.50-4161.00\AA & \citet{Kauffmann_etal_2003} \\
H$\alpha$ & 6554.60-6574.60\AA &
  6483.00-6513.00\AA & 6623.00-6653.00\AA & \citet{Yan_etal_2006} \\
\hline
\end{tabular}

\caption{The definition of the wavelength ranges we used to calculate the
equivalent widths for different lines. A line is fitted through both the blue
and red sidebands to represent continuum level, which is used for the
equivalent width calculation in the central bandpass. The references for the
use of these exact values are given in the last column.}

\label{tab:ew_wavelength}
\end{center}
\end{table*}

Finally, after the fitting process and various cuts, we visually inspect the
spectra of all candidate post-starburst galaxies. There is a fairly high false
positive rate (objects that are flagged by the selection process as candidates,
but turn out to be other classes of objects). Most of these are
catastrophically wrong fits in which the templates happen to match to certain
spectral features producing a (small) maximum value of $\Delta\chi^2$ at some
assumed redshift while at the same time giving $\mathrm{A/Total}>0.25$. These
objects are removed from the candidate list at this stage. The other class of
contamination is A-stars and white dwarfs whose spectra fit the ``A''-template
extremely well at $z=0$ by construction. This problem only exists for BOSS
spectra because the wavelength coverage extends blue enough for the 3600-4400
\AA\, range to be within the data coverage at zero redshift. We solve this
issue by requiring the best-fit redshift to be bigger than $z = 0.02$, which
vastly reduces the number of candidates to be visually inspected. The numbers
of objects considered at each stage in the selection process are shown in Table
\ref{tab:number_selection}. This selection method does a decent job of
screening the entire spectroscopic database and selecting a relatively small
number of potential candidates. Yet it still has a high false positive rate,
about 300\% for SDSS and 100\% for BOSS, requiring human inspection of the
spectra.

\begin{table}
\begin{center}
\begin{tabular}{|c|c|c|}
\hline
Step & SDSS & BOSS \\
\hline
All Spectra & $\sim 1.7 \times 10^6$ & $\sim 7.6 \times 10^5$ \\
Program Candidates & 8216 & 23728 \\
Program Candidates ($z>0.02$) & 8216 & 8485 \\
Visually Inspected & 2330 & 3964 \\
\hline
\end{tabular}

\caption{The number of spectra considered at each stage of the selection
process. The first row is the total number of spectra considered, essentially
every spectrum available in the SDSS DR9. The second row is the number of
spectra flagged by the selection method to be possible post-starburst galaxy
candidates after the fitting process and the application of the cuts on the
A/Total ratio and the equivalent widths of the H$\delta$ and [OII]$\lambda$3727
lines. The third row is the same as the second row but with the requirement
that $z > 0.02$ to remove the contamination from A-stars and white dwarfs. This
redshift restriction does not affect SDSS sample since the wavelength coverage
does not allow us to probe to such low redshift. The last row is the number of
post-starburst galaxies in the final sample after the visual inspection.}

\label{tab:number_selection}
\end{center}
\end{table}

This selection method recovers most of the known post-starburst galaxies
described in Section \ref{sec:known_sample}. More specifically, 452 objects out
of 564 objects ($\sim$80\%) in the \citet{Goto_2007} sample are recovered. The
ones that are not recovered are visually confirmed to either have
signal-to-noise ratio too low or are simply contaminants to that sample.
Therefore we believe that this selection method is robust and is able to
identify most post-starburst galaxies in the sample, with a low false negative
rate. It is not surprising that this selection method is robust when used
against this known sample, since it was designed around this very sample.
However, since the method is generic and the amount of fine-tuning is minimal,
we expect that this selection method is robust in general.

\section{Sample Properties}
\label{sec:properties}

In this section we present our final sample, which is distributed uniformly
over the survey footprint as expected. An {\tt ASCII} file containing the full
list of post-starburst galaxies presented in this paper, along with important
properties, is included. Table \ref{tab:sample_table} shows example entries of
that full list.

\begin{figure}[h]
\begin{center}
\includegraphics[trim=20 5 10 20,clip,width=0.475\textwidth]{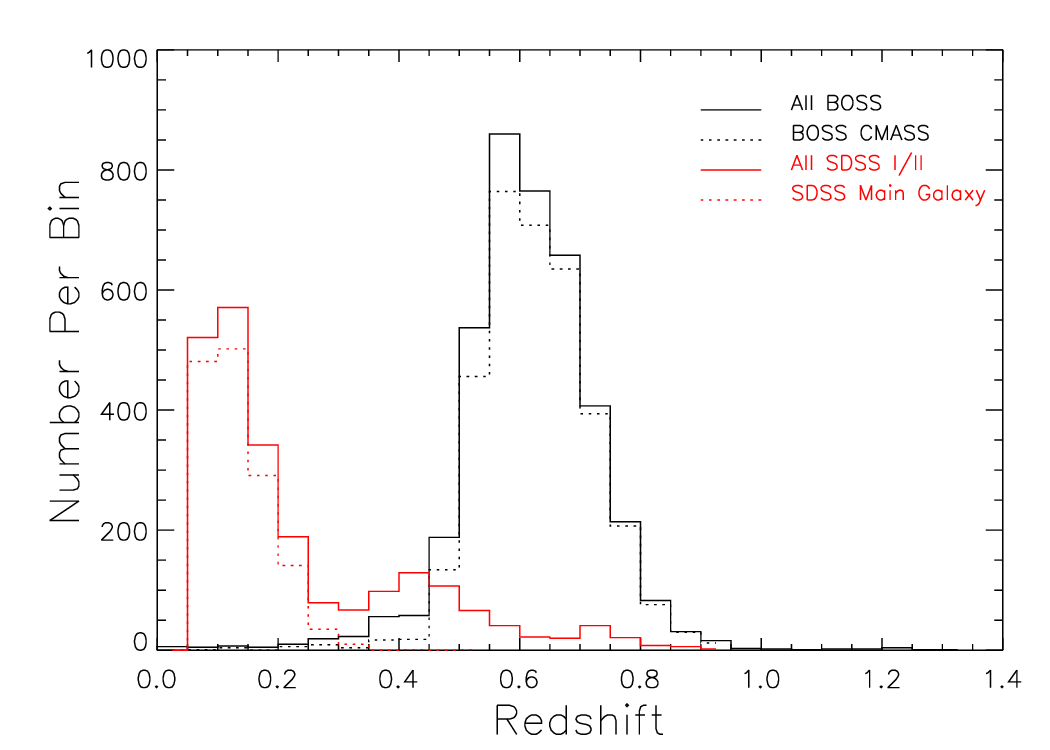}

\caption{Redshift distributions of post-starburst galaxy samples from both SDSS
I/II and BOSS. The solid lines of different colors (red: SDSS I/II, black:
BOSS) show the distributions of the entire respective samples, while the dotted
lines show the subsamples used to calculate the luminosity function (the Main
Galaxy Sample for SDSS, and the CMASS Galaxy Sample for BOSS). Note the tail of
objects with redshifts up to $z\sim1.3$.}

\label{fig:redshift_histogram}
\end{center}
\end{figure}

\begin{figure*}
\begin{center}
\includegraphics[trim=55 5 30 10,clip,width=\textwidth]{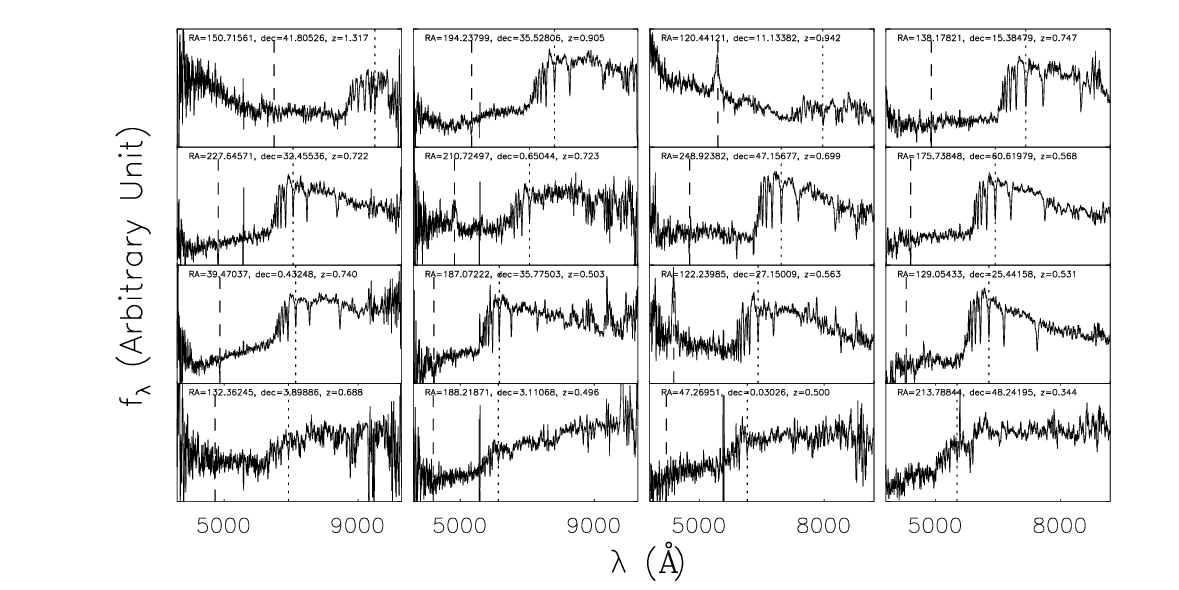}

\caption{Spectra of a number of objects selected from our final sample. The
left half shows objects from the BOSS sample while the right half shows ones
from the SDSS I/II sample. The top left corner panel of each side shows the
highest-redshift object in each sample. The bottom four panels show objects
with noisy spectra to demonstrate the lowest signal-to-noise ratio that could
be firmly identified as post-starburst galaxies. The rest of the panels are
selected to roughly represent the range of redshift and A/Total ratio spanned
by the sample and to also show objects with interesting spectral features, such
as blue continuum or MgII$\lambda$2803 line in either absorption and emission.
All spectra are smoothed by 10 pixels to show features clearly. Dashed and
dotted vertical lines show the position of the MgII$\lambda$2803 and H$\delta$
lines respectively. Many spectra show a residual of the strong sky line at
5577\AA.}

\label{fig:spectra_sample}
\end{center}
\end{figure*}

\begin{figure}[h]
\begin{center}
\includegraphics[trim=15 0 15 10,clip,width=0.475\textwidth]{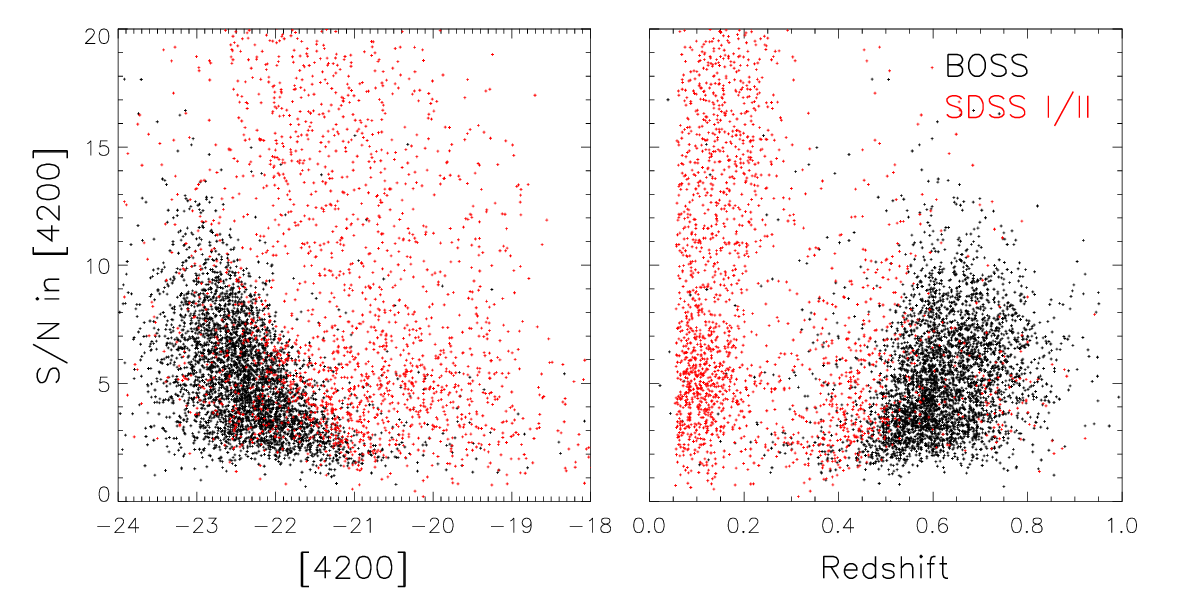}

\caption{Median signal-to-noise ratio per pixel in the rest-frame wavelength
range 4150-4300\AA\, (the ``[4200] band'') of post-starburst galaxies in our
sample, shown as a function of the rest-frame absolute magnitude in the same
band (see Section \ref{sec:fid_band} for more details) and redshift. The red
and black points represent objects from SDSS-I/II and SDSS-III BOSS samples
respectively. There are very few objects with S/N$<2$, which is an empirical
lower limit for robustly identifying post-starburst galaxies.}

\label{fig:signal_noise}
\end{center}
\end{figure}

The redshift distributions of the post-starburst galaxies selected from both
SDSS and BOSS are shown in Figure \ref{fig:redshift_histogram}. The median
redshifts are around $0.2$ for SDSS and $0.6$ for BOSS respectively, with very
little overlap between the two samples. Example of the spectra of these objects
are shown in Figure \ref{fig:spectra_sample}. These examples are selected to
illustrate a range of properties spanned by the objects in our sample. A
diverse range of spectral features are seen among the objects in this sample.
First is the strong Balmer lines, especially H$\delta$, in absorption which is
one of the requirements for the selection. A number of objects show a blue
continuum blueward of 4000\AA, which is a sign of either a young stellar
population or AGN activity. A large number of objects show weak MgII
absorption, some even with detectable blueshift, indicative of outflows, which
may be driven by either an AGN or a compact starburst
\citep{Tremonti_etal_2007, Diamond_Stanic_etal_2012, Sell_etal_2014}. More rare
are objects with broad MgII in emission, clearly indicative of AGN activity.
These are ``post-starburst quasars'' (for example, \citet{Brotherton_etal_1999,
Brotherton_etal_2004}). The H$\alpha$ line for lower-redshift objects also
shows a diverse range of behaviors, ranging from absorption to broad line
emission. This H$\alpha$ behavior and how it is related to the contamination in
the selection will be discussed in Section \ref{sec:halpha_ew}. The quantities
A/Total and H$\delta$ absorption Equivalent Width are roughly correlated. This
is expected since both are indicators of the strength of the A-star component.
Figure \ref{fig:signal_noise} shows the distribution of the signal-to-noise
ratio in the 4150-4300\AA\, range (median of the per pixel signal-to-noise
ratios for all pixels in this wavelength range) of the sample as a function of
redshift and rest-frame absolute magnitude in the same wavelength range
(denoted by [4200]; for more details on the calculation of this magnitude, see
Section \ref{sec:fid_band}) respectively. Even though we have not made explicit
cuts in signal-to-noise ratio, 96\% of the objects have signal-to-noise ratio
larger than 2, an approximate lower limit to robustly identify post-starburst
galaxies (Figure \ref{fig:signal_noise}). Therefore, the mis-identification due
to low signal-to-noise ratio is not a major source of concern for this sample.
We quantify the effects of signal-to-noise ratio on selection further below.

\begin{table*}
\small
\begin{center}
\begin{tabular}{|c|c|c|c|c|c|c|c|c|c|c|c|}
\hline
Survey & RA($^\circ$) & DEC($^\circ$) & Plate & Fiber & MJD & $z$ & H$\delta$EW & [OII]$\lambda$3727EW & A/Total & Target Selection & $z=z_\mathrm{pipe}$ \\
\hline
SDSS & 120.44121 & 11.133820 & 2418 & 499 & 53794 & 0.942 & 4.80 & -1.75 & 0.32 &       36700160 & False  \\
SDSS & 247.71479 & 47.069620 &  627 & 377 & 52144 & 0.825 & 8.06 &  5.80 & 0.76 &       33554433 & True   \\
SDSS & 35.161773 &0.53262315 & 2637 & 205 & 54504 & 0.646 & 8.65 & -0.29 & 0.67 &             32 & True   \\
BOSS & 165.43852 & 37.536575 & 4626 & 730 & 55647 & 1.292 & 7.43 & -0.26 & 0.54 & 10995116687360 & True   \\
BOSS & 233.62838 &0.72746109 & 4010 & 118 & 55350 & 0.941 & 8.75 & -2.00 & 0.34 &  3298535538688 & True   \\
BOSS & 124.34978 & 34.166264 & 3758 & 865 & 55506 & 0.823 & 8.41 &  0.45 & 0.69 &            134 & True   \\
\hline
\end{tabular}

\caption{Example of the full list of post-starburst galaxies presented in this
paper to demonstrate the format. The first column indicates whether the object
belongs to the SDSS I/II or BOSS subsample. Columns 2-6 are for identification:
Right Ascension, Declination and the Plate/Fiber/MJD identification of SDSS
spectroscopy. Column 7 is the redshift determined from our selection method.
Columns 8-10 are properties used in the selection method, namely the equivalent
widths of H$\delta$ and [OII]$\lambda$3727 lines, and the A/Total ratio from
template fit. Column 11 is the bitmask showing the spectroscopic target
selection algorithm. The bitmask can be decoded using information available on
the SDSS website {\tt
http://www.sdss3.org/dr10/algorithms/bitmask\_legacy\_target1.php} for SDSS
I/II objects, and {\tt
http://www.sdss3.org/dr10/algorithms/bitmask\_boss\_target1.php} for BOSS
objects. The last column indicates whether the pipeline redshift
\citep{Bolton_etal_2012} agrees with the fit redshift. The objects in this
table are unique objects, separated by SDSS I/II and BOSS surveys, and ordered
by redshift. The full table is available in {\tt ASCII} format.}

\label{tab:sample_table}
\end{center}
\end{table*}

\subsection{Coadded Spectra}

\begin{figure}
\begin{center}
\includegraphics[trim=25 5 5 20,clip,width=0.475\textwidth]{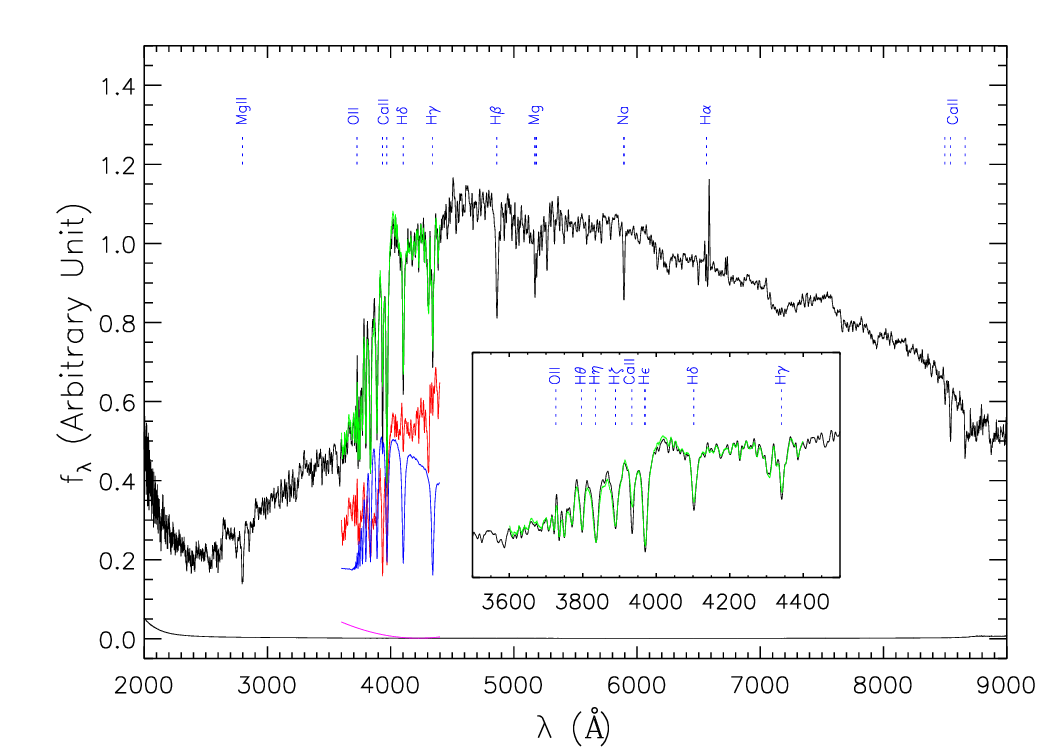}

\caption{The coadded spectrum of the post-starburst galaxy sample from both
SDSS and BOSS, with line features identified. The black solid line at the
bottom of the plot shows the per-pixel error of the spectrum. The errors are
very close to zero over most of the spectral range, indicative of very high
signal-to-noise ratio. The inset shows the spectrum in the 3600-4400 \AA\,
region, which is used for template fitting, in more detail. The green curve
shows the model fit to this coadded spectrum, with the ``K'', ``A'' and
Polynomial components of the model shown in red, blue and magenta respectively.
Note that the polynomial component is small in this coadded spectrum, as it is
for most individual objects. It should be noted that at the short and long
wavelength ends ($\lambda<3000$\AA\, and $\lambda>8000$\AA), the coadded
spectrum is derived from only a small number of objects and thus may not be
very reliable.}

\label{fig:stacked_spectrum}
\end{center}
\end{figure}

The spectra from the sample were coadded to explore post-starburst features at
very high signal-to-noise ratio. To do this, the spectra are summed with
inverse variance weighting, after bringing all spectra to zero redshift and
scaling each by the mean flux density between 4150-4300 \AA. The final coadded
spectrum, along with the fit of this spectrum into components (Equation
\ref{eq:model_fit}), are shown in Figure \ref{fig:stacked_spectrum}. This
combined spectrum calculated from the whole sample of several thousand galaxies
has very high signal-to-noise ratio and many spectral features are identified
with high confidence. A small amount of [OII]$\lambda$3727 and H$\alpha$ in
emission can be seen under closer inspection. This is due to emission in
individual objects that are small enough to pass the equivalent width cuts.

The coaddition is also done for various different subgroups of galaxies in
order to investigate whether the spectra change systematically with properties
of the galaxies. The coadded spectra of objects in two redshift bins are in
excellent agreement, showing that most of the small-scale features are real. On
the other hand, the continuum colors of two redshift bins differ somewhat. As
most objects in the low-redshift bin are from SDSS I/II while the high redshift
bin is dominated by BOSS objects, this reflects the different selection of the
two samples. This selection effect is properly corrected for in the luminosity
function analysis in Section \ref{sec:luminosity_function}.

\begin{figure}
\begin{center}
\includegraphics[trim=20 0 5 15,clip,width=0.475\textwidth]{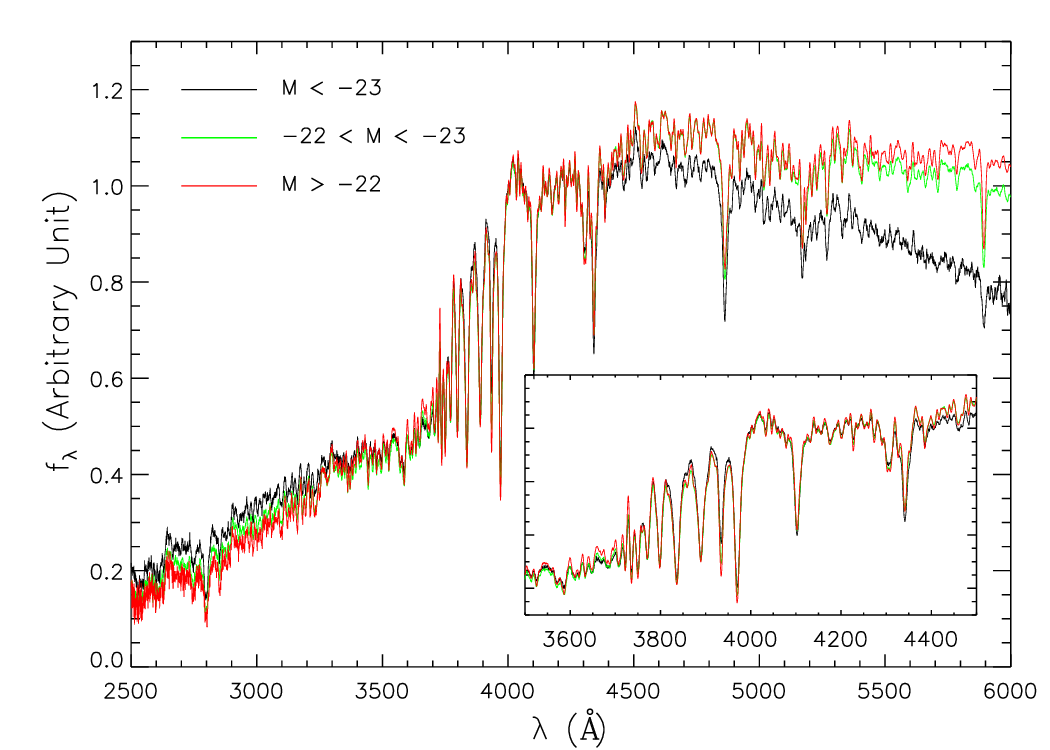}

\caption{The coadded spectrum of the post-starburst galaxy sample binned by
absolute magnitude in our fiducial [4200] band. The black, green and red lines
are the most, intermediate and least luminous groups respectively. The small
spectral features match in detail in these independent coadds, demonstrating
that they are real. The differences in overall SED shapes may be due to either
the intrinsic properties of the stellar populations, or selection effects.}

\label{fig:stacked_spectrum_luminosity}
\end{center}
\end{figure}

Figure \ref{fig:stacked_spectrum_luminosity} shows the coadded spectra binned
by absolute magnitude in the fiducial band at 4200\AA. Again, the fine features
in the spectra match extremely well between the different bins. The highest
luminosity group of objects are significantly bluer longward of $4500$\AA.
There is also a slight difference in continuum level blueward of $3500$\AA,
where the highest-luminosity galaxies have slightly higher continuum level. It
is not clear if these differences are due to astrophysical or selection
effects. If they are astrophysical, then it means that the more intrinsically
luminous objects are more dominated by the A-star population, which has a bluer
continuum. On the other hand, these differences could be due to selection
effects, because more luminous objects tend to be in the BOSS CMASS Sample,
while less luminous objects are likely to be from the SDSS Main Galaxy Sample.

\begin{figure}
\begin{center}
\includegraphics[trim=20 0 5 15,clip,width=0.475\textwidth]{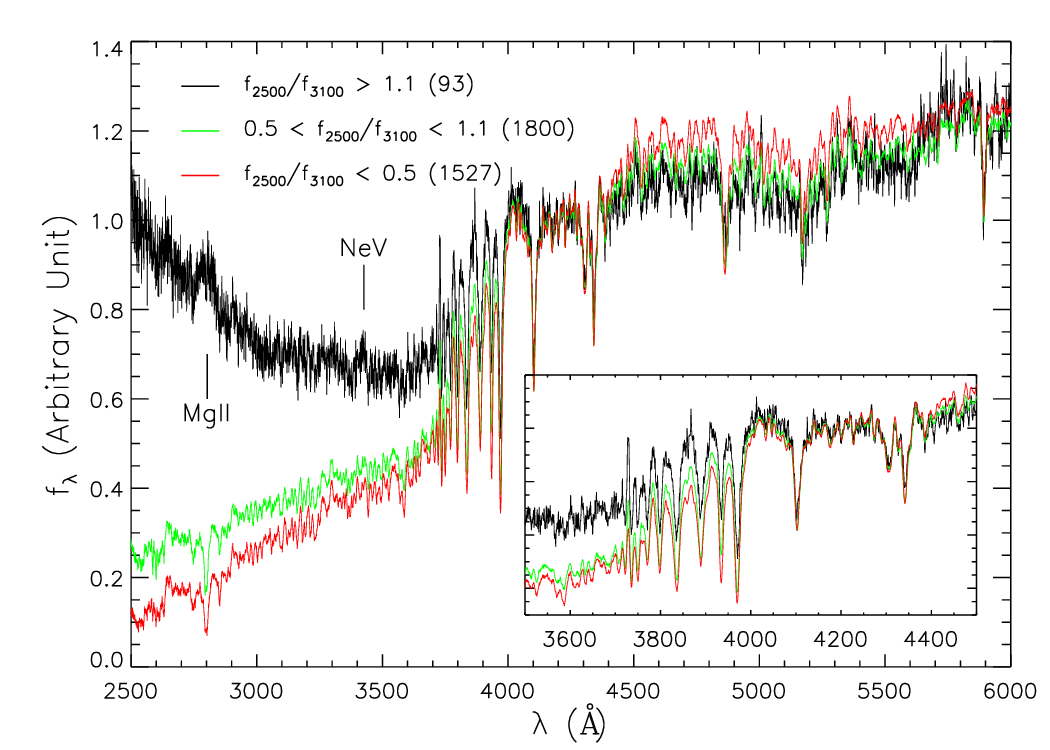}

\caption{The coadded spectra of the post-starburst galaxy sample binned by the
continuum color below $3500$\AA, defined as the ratio of flux density at
$2500$\AA\, to that at $3100$\AA, averaged over bandpasses with $200$\AA\,
width. The numbers of objects in each bin are shown in parentheses. The most
dramatic feature, in addition to the slight difference in continuum color at
the red side of $4000$\AA\, break, is that the composite spectrum of objects
with extremely blue continuum shows broad line MgII$\lambda$2803 emission,
while the rest of the sample shows it in absorption. Additionally,
NeV$\lambda$3426 in emission can be seen in this population. That is, these are
post-starburst quasars, showing both AGN features (blue continuum, broad line
emission) and post-starburst spectral features.}

\label{fig:stacked_spectrum_bluecon}
\end{center}
\end{figure}

Figure \ref{fig:stacked_spectrum_bluecon} shows the coadded spectra binned by
the continuum color below $3500$\AA, defined as the ratio of the flux density
at $2500$\AA\, to that at $3100$\AA, averaged over bandpasses with $200$\AA\,
width. The composite of the small number of objects with very blue continua (93
out of $\sim3500$ which have high enough quality spectra and lie at high enough
redshift that this exercise can be carried out) have broad line emission in
MgII$\lambda$2803 and NeV$\lambda$3426, while the rest of the sample tends to
have MgII in absorption. This broad line emission is the distinct signature of
post-starburst quasars. Objects with blue continua also tend to show stronger
[OII]$\lambda$3727 line in emission than the rest of the sample. The difference
is small by construction since any object with stronger [OII] emission would
not be selected into the sample. This increased [OII] emission can be due to
either a small amount of ongoing star formation, or more likely, AGN activity
as already suggested by broad MgII emission. This sample can be compared to a
bigger sample of SDSS post-starburst quasars presented by
\citet{Brotherton_etal_2004}. Most of those objects have strong enough [OII]
emission that they would be excluded from our sample by our requirement on
[OII] equivalent width.

\subsection{Spectroscopic Target Selection}
\label{sec:target_selection}

SDSS and BOSS choose targets for spectroscopic observation based on colors and
morphology from the imaging data. Each spectroscopic target is selected by
(possibly multiple) selection criteria. Therefore, in order to understand the
properties of this sample, it is crucial to understand how each object is
selected in the spectroscopic sample into the first place.

Tables \ref{tab:sdss_target_flag} and \ref{tab:boss_target_flag} show the
breakdown of target selection algorithms by which our post-starburst galaxies
are selected, from the SDSS I/II and BOSS sample respectively. In the SDSS I/II
sample, the {\tt GALAXY} flag is the dominant target selection algorithm. This
corresponds to the Main Galaxy Sample, which is the $r_\mathrm{petro} < 17.77$
magnitude-limited galaxy sample described in \citet{Strauss_etal_2002}.

The only other dominant selection flag among the SDSS I/II sample is {\tt
QSO\_HIZ}, which is designed for the high-redshift quasars. A significant
number of objects are selected through this flag because they appear
unresolved, as required by the high-redshift quasar selection criteria, and
the broadband color of the Balmer break at redshift less than unity resembles
the Lyman break at redshift $z \sim 3-4$ \citep{Fan_etal_1998}.

For the BOSS sample, the dominant selection flags are all related to the CMASS
program, which is the ``Constant Stellar Mass'' sample of luminous galaxies
with redshift $0.4 < z < 0.7$ designed to measure Baryon Acoustic Oscillations
\citep{Anderson_etal_2012, Dawson_etal_2013}. The most important of these flags
is the {\tt GAL\_CMASS} flag. The flags {\tt GAL\_CMASS\_COMM} and {\tt
GAL\_CMASS\_ALL} were used in the early part of the survey to tune the CMASS
selection criteria, and are therefore more inclusive but less complete. When we
calculate the luminosity function for the CMASS galaxies (Section
\ref{sec:luminosity_function}), we limit our sample to the objects selected by
the {\tt GAL\_CMASS} flag only. It should also be noted that very few objects
in our sample are part of the other major BOSS galaxy sample, the LOWZ sample.

\subsection{Comparison to the Pipeline Redshifts}
\label{sec:compare_pipeline_redshift}

We mentioned in Section \ref{sec:redshift_determination} that some of the
high-redshift post-starburst galaxies have incorrect pipeline redshifts. We now
discuss this issue in more detail.

Table \ref{tab:pipeline_z_diff} summarizes the comparison between pipeline
redshifts and our measured redshifts for our post-starburst galaxy sample. The
sample from SDSS I/II is broken down by availability of the second pipeline
redshift. For this sample, the main spectroscopic pipeline ({\tt idlspec2d})
redshift \citep{Bolton_etal_2012} and the {\tt spectro1d} redshift
\citep{SubbaRao_etal_2002} are each incorrect for 4-6\% of our galaxies. The
BOSS post-starburst sample has better redshift determination; it is incorrect
for only about 2\% of our objects. The main reason the pipeline redshifts are
incorrect is the lack of post-starburst spectral templates in the spectroscopic
pipeline. This was improved in the BOSS sample, resulting in better redshift
determination. Other reasons for incorrect redshift determinations are
artifacts in the data that significantly change the shape of the continuum,
such as bad CCD columns or light leakage from bright object in adjacent fibers,
or simply too low signal-to-noise ratio. In most (but not every) of these
cases, the {\tt ZWARNING} flag is set by the pipeline to indicate low
confidence in the redshift.

\begin{figure*}
\begin{center}
\includegraphics[trim=10 0 25 15,clip,width=\textwidth]{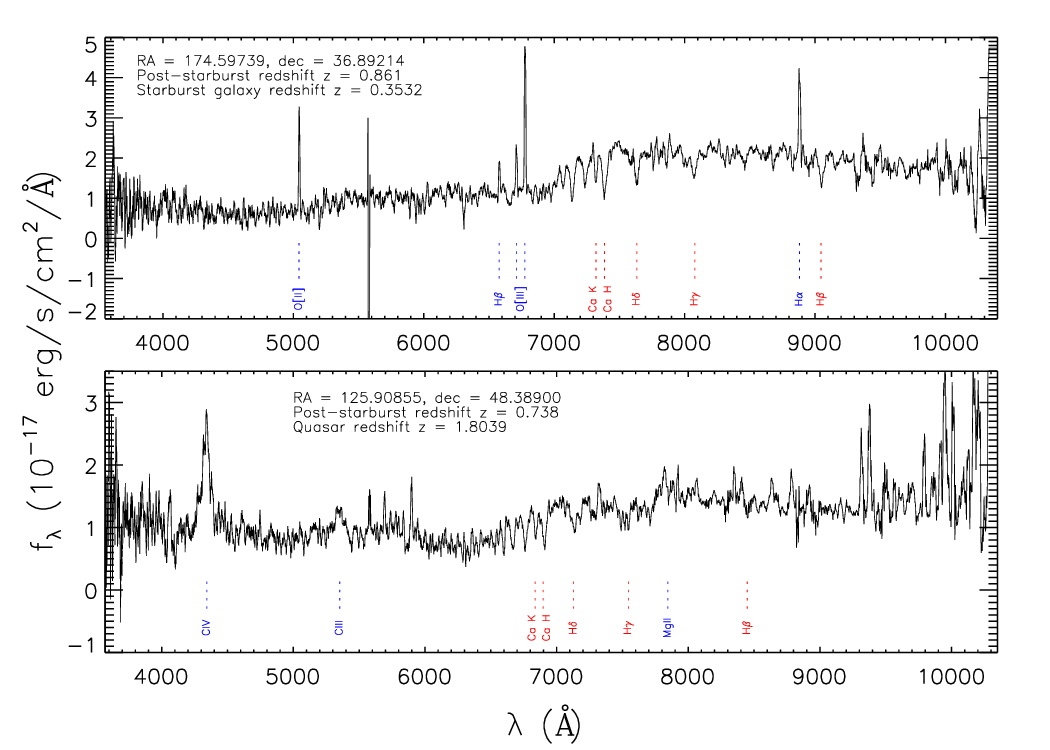}

\caption{Two interesting objects that were discovered serendipitously because
the pipeline redshifts and the fit redshifts disagree. They are actually pairs
of objects along the same line of sight at different redshifts. The top panel
shows a post-starburst galaxy at redshift $z=0.861$ with a star-forming galaxy
at redshift $z=0.3532$. The bottom panel is a post-starburst galaxy at redshift
$z=0.738$ with a quasar at redshift $z=1.8039$. The spectral features belonging
to the post-starburst galaxies in each plot are indicated in red, while the
ones belonging to the superposed objects are indicated in blue. These spectra
are smoothed with a 10 pixel boxcar average to show the features clearly.}

\label{fig:interesting_spectra}
\end{center}
\end{figure*}

Two objects with discrepant redshifts are particularly interesting: they are
actually pairs of objects along the same line of sight at different redshifts.
In these cases, our post-starburst galaxy selection method identifies one
object of the pair, while the SDSS spectroscopic pipeline correctly identifies
another. These systems are shown in Figure \ref{fig:interesting_spectra} with
various line features identified. The first system is a post-starburst galaxy
at redshift $z=0.861$ with a star-forming galaxy at redshift $z=0.3532$. The
second system is a post-starburst galaxy at redshift $z=0.738$ and a quasar at
redshift $z=1.8039$. These objects may be interesting for follow-up
observations to study strong gravitational lensing or absorption spectroscopy
of outflows, for example.

\begin{table}
\begin{center}
\begin{tabular}{|c|c|}

\hline
Sample                      &   Number of Objects   \\
\hline

\hline
BOSS entire sample                                      &   $3964$              \\
$z_\mathrm{fit} = z_1$                                  &   $3900 = 98.38\%$    \\
\hline

\hline
SDSS entire sample                                      &   $2330$              \\
$z_\mathrm{fit} = z_1$                                  &   $2194 = 94.16\%$    \\
\hline

\hline
SDSS $z_2$ exists                                       &   $2223$              \\
$z_\mathrm{fit} = z_1$                                  &   $2106 = 94.74\%$    \\
$z_\mathrm{fit} = z_2$                                  &   $2153 = 96.85\%$    \\
$z_\mathrm{fit} = z_1$ but $z_\mathrm{fit} \neq z_2$    &   $47 = 2.11\%$       \\
$z_\mathrm{fit} = z_2$ but $z_\mathrm{fit} \neq z_1$    &   $94 = 4.23\%$       \\
$z_\mathrm{fit} = z_1 = z_2$                            &   $2059 = 92.62\%$    \\
$z_\mathrm{fit} \neq z_1 \neq z_2$                      &   $23 = 1.03\%$        \\
\hline

\hline
SDSS $z_2$ does not exist                               &   $107$               \\
$z_\mathrm{fit} = z_1$                                  &   $88 = 82.24\%$      \\
\hline

\end{tabular}

\caption{The breakdown of our post-starburst galaxy sample in terms of whether
the redshifts determined by the spectroscopic pipelines agree with those from
our fitting procedure. In this table, $z_1$ indicates the {\tt idlspec2d}
redshift from the main spectroscopic pipeline \citep{Bolton_etal_2012} which is
available for both samples, while $z_2$ is the {\tt spectro1d} redshift
\citep{SubbaRao_etal_2002}, which is available for most of the SDSS I/II
sample, but not for BOSS.}

\label{tab:pipeline_z_diff}
\end{center}
\end{table}

\begin{table}
\begin{center}
\begin{tabular}{|c|c|}
\hline
Selection Criteria          & Number        \\
\hline
Total                       & 2330          \\
Main Galaxy Sample          & 1462          \\
All Galaxy                  & 1879          \\
Luminous Red Galaxy         & 140           \\
Quasar                      & 406           \\
ROSAT                       & 32            \\
Stellar                     & 30            \\
Serendipitous               & 66            \\
\hline
\end{tabular}

\caption{The breakdown of our post-starburst galaxies sample from the SDSS I/II
sample in terms of the spectroscopic target selection algorithms by which they
are selected. The second row is the magnitude-limited main galaxy sample that
we use in our luminosity function analysis. The other rows show the number of
objects selected by various broad groups of selection algorithms. Note that a
galaxy can be selected by more than one algorithm.}

\label{tab:sdss_target_flag}
\end{center}
\end{table}

\begin{table}
\begin{center}
\begin{tabular}{|c|c|}
\hline
Selection Criteria          & Number        \\
\hline
Total                       & 3964          \\
CMASS Sample                & 3475          \\
Galaxy                      & 3810          \\
Quasar                      & 72            \\
Spectral Template           & 5             \\
\hline
\end{tabular}

\caption{The breakdown of our post-starburst galaxies sample from the BOSS
sample in terms of the spectroscopic target selection algorithms by which they
are selected. The second row is the CMASS Sample with well-defined selection
that we use in our luminosity function analysis. The other rows show the number
of objects selected by various broad groups of selection algorithms. Note that
a galaxy can be selected by more than one algorithm.}

\label{tab:boss_target_flag}
\end{center}
\end{table}

\subsection{H$\alpha$ Equivalent Width}
\label{sec:halpha_ew}

In our selection of this sample, we have neglected the H$\alpha$ line and only
considered two lines (H$\delta$ in absorption and lack of [OII]$\lambda$3727 in
emission) since most objects we select for are at high enough redshift that the
H$\alpha$ line is redshifted out of the wavelength coverage of the spectra. In
order to compare this to other works and understand any biases the lack of
selection in H$\alpha$ might introduce, we need to quantify how strong
H$\alpha$ is in these objects. This exercise is important for two reasons. The
first is because the H$\alpha$ line is commonly used as an indicator of star
formation rate in most previous studies of E+A galaxies at low redshift. The
second reason is that AGN are also known to produce H$\alpha$ emission, and
therefore can introduce selection biases. We address this concern with two
different approaches.

\begin{figure}
\begin{center}
\includegraphics[trim=15 5 10 15,clip,width=0.475\textwidth]{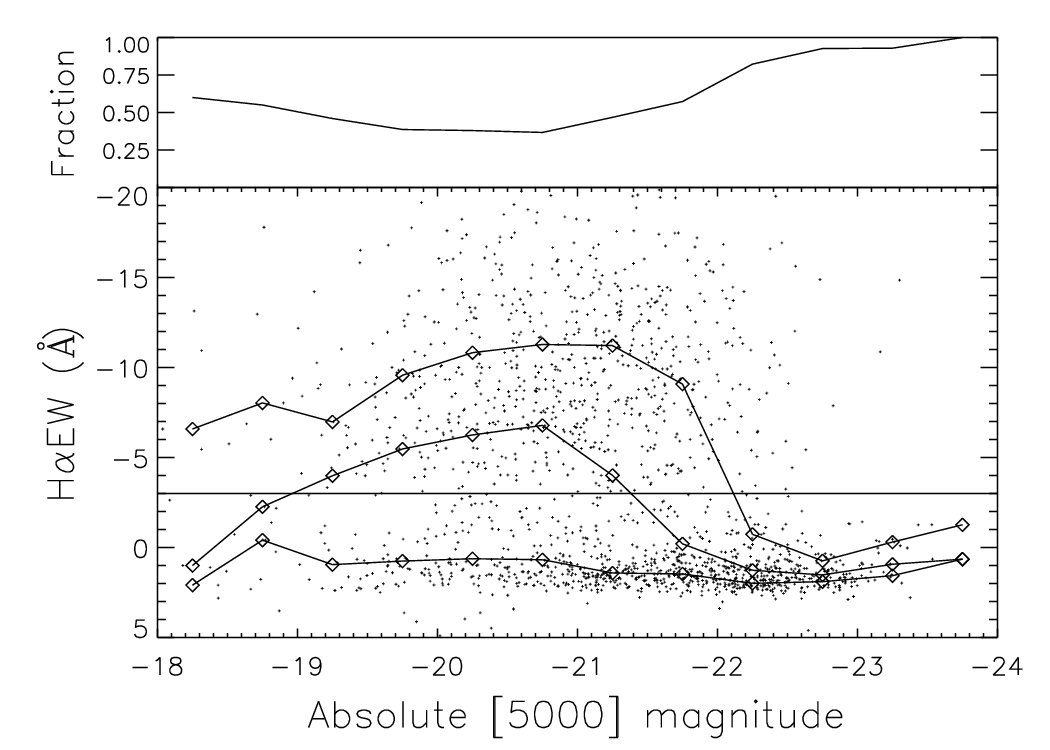}

\caption{Rest-frame H$\alpha$ Equivalent Width against the absolute magnitude
in the [5000] band for post-starburst galaxies from SDSS I/II sample with
redshift $z<0.33$, selected irrespective of their H$\alpha$ properties. In the
sign convention used here, positive values of Equivalent Width correspond to
absorption. The horizontal line indicates the selection criteria H$\alpha$EW
$>$ $-3.0$\AA\, used by \citet{Goto_2007}. The three lines with square symbols
are the 25, 50 and 75 percentiles of the objects in each half-magnitude bin
respectively. The top panel shows the fraction of objects that pass the
\citet{Goto_2007} selection criteria in each bin.}

\label{fig:halpha_luminosity}
\end{center}
\end{figure}

The first one is to use our sample (mainly from SDSS-I/II) at low enough
redshift that H$\alpha$ lies in the SDSS spectra. Figure
\ref{fig:halpha_luminosity} shows the distribution of H$\alpha$ equivalent
width as a function of the absolute magnitude in the rest-frame [5000] band
(this band is used in Section \ref{sec:fid_band} for our luminosity function
calculation) for the 1766 SDSS-I/II post-starburst galaxies at $z < 0.33$.
Among these objects only 968 (54\%) have H$\alpha$EW $>$ $-3.0$\AA, the
selection criterion adopted by \citet{Goto_2007}. This means that nearly half
of the sample at low redshift would be dropped as star forming if H$\alpha$
were taken into consideration, a rate consistent with what \citet{Goto_2007}
found. However, if one limits the consideration to only the intrinsically
luminous galaxies, the contamination is minimal. This can be seen from the
subsample with absolute [5000] magnitude brighter than $-22.5$. Out of 154
objects in this subsample, only 11 objects (7\%) have H$\alpha$EW $>$
$-3.0$\AA. Since most of our targets at high redshifts selected from BOSS are
more luminous than $M\sim-22.5$ in the same band, and under the assumption that
the post-starburst galaxies of the same intrinsic brightness have the same
properties across cosmic time, we can extrapolate from this bright,
low-redshift sample to conclude that our high-redshift sample from BOSS is
minimally contaminated by our lack of H$\alpha$ information.

At the high-redshift end, we conducted follow-up near-IR spectroscopic
observations on a small number of objects to observe H$\alpha$ directly. We
observed 10 objects at $z \sim 0.8$ with the TripleSpec instrument
\citep{Wilson_etal_2004} on the ARC 3.5m telescope at the Apache Point
Observatory. The follow-up observations took place between November 2011 and
November 2013, with details shown in Table \ref{tab:tspec_info}. The targets
were selected such that the H$\alpha$ line is expected to be in a region clear
of telluric absorption, and to be as bright as possible for maximum
signal-to-noise ratio. The exposure times were between 40 minutes and 104
minutes per object. The spectra were reduced using the publicly available
software TripleSpecTool, which is a modified version of Spextool developed by
\citet{Cushing_Vacca_Rayner_2004}. The software uses the telluric correction
algorithm developed by \citet{Vacca_Cushing_Rayner_2003}.

\begin{figure}
\begin{center}
\includegraphics[trim=40 10 20 20,clip,width=0.475\textwidth]{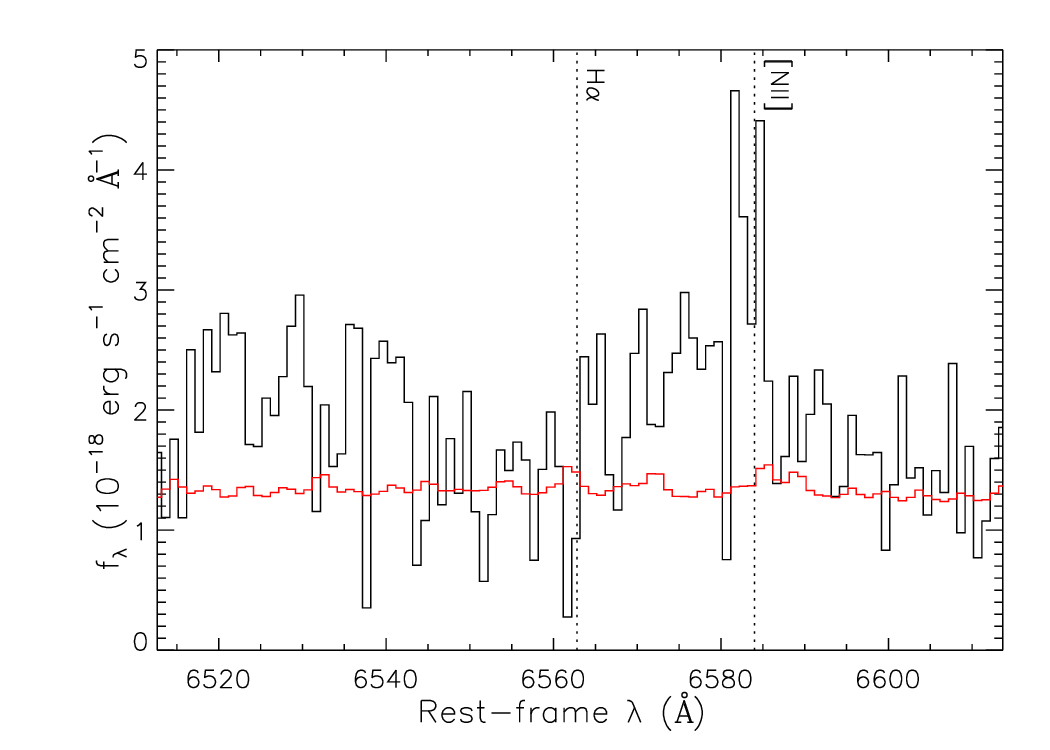}

\caption{The composite spectrum around the wavelength of H$\alpha$ of 10
post-starburst galaxies that were observed by the TripleSpec instrument at the
APO 3.5m telescope. The noise level is shown by the red line. The expected
positions of H$\alpha$ and [NII]$\lambda$6584 are shown by the dotted vertical
lines. The continuum of these objects is clearly detected, but the
signal-to-noise ratio is too low for H$\alpha$ to be seen either in emission or
absorption. The [NII] emission is detected with reasonable confidence.}

\label{fig:tspec_stacked}
\end{center}
\end{figure}

The objects are clearly detected in the continuum, but are too faint to have
good enough signal-to-noise ratio to detect the H$\alpha$ line either in
absorption or emission. The composite spectrum of these objects at rest-frame
H$\alpha$ wavelength is shown in Figure \ref{fig:tspec_stacked} and the
equivalent width of H$\alpha$ calculated from this composite using a simple
estimate is $\mathrm{H}\alpha\mathrm{EW}\approx-4\pm6$ \AA, consistent with
zero. This lack of any clear detection is consistent with our low-redshift
finding that luminous post-starburst galaxies selected only from H$\delta$ and
[OII]$\lambda$3727 are minimally contaminated by objects with large H$\alpha$
emission. The [NII]$\lambda$6584 emission is detected in this composite
spectrum with reasonable confidence. At face value, high [NII] to H$\alpha$
ratio suggests that the spectrum is dominated by an AGN. However, it is also
possible that the H$\alpha$ absorption from the post-starburst spectrum is
offset by H$\alpha$ emission from a starburst or AGN component resulting in a
non-detection in this line (e.g., \citet{Rubin_Ford_1986}).

\begin{table}
\begin{center}
\begin{tabular}{|c|c|c|c|c|}
\hline
Date observed &     RA      &       DEC    &  Redshift & Exposure time \\
\hline
 16 Nov 2011  & 01:36:11.85 & +00:30:04.39 &  0.7886   &    48 mins    \\
 16 Nov 2011  & 02:49:56.15 & -00:59:05.65 &  0.8270   &    56 mins    \\
 02 Jun 2012  & 16:46:24.77 & +28:23:12.56 &  0.8560   &    64 mins    \\
 12 Jun 2012  & 12:32:54.27 & +00:22:43.12 &  0.8499   &    40 mins    \\
 05 Sep 2013  & 21:15:33.74 & +00:22:47.93 &  0.8338   &    84 mins    \\
 05 Sep 2013  & 01:13:33.06 & +02:50:26.32 &  0.8980   &    92 mins    \\
 05 Sep 2013  & 02:29:24.93 & -03:36:33.60 &  0.8520   &    56 mins    \\
 19 Nov 2013  & 21:20:56.53 & +00:01:13.24 &  0.8381   &    76 mins    \\
 19 Nov 2013  & 23:51:41.07 & +03:05:42.60 &  0.8570   &    60 mins    \\
 19 Nov 2013  & 02:08:51.52 & +05:52:28.04 &  0.8269   &   104 mins    \\
\hline
\end{tabular}
\end{center}

\caption{Details of follow-up observations of high-redshift post-starburst
galaxies using the TripleSpec instrument at the Apache Point Observatory (APO)
3.5m telescope.}

\label{tab:tspec_info}
\end{table}

\subsection{Potential Biases in Selection}
\label{sec:selection_bias}

In this section, we discuss potential selection biases in our post-starburst
galaxy sample. It is common to use the strength of [OII]$\lambda$3727 and
H$\alpha$ emission to measure the star formation rate, in order to make cuts in
these quantities and select objects with little or no ongoing star formation.
However, AGN also produce emission in both lines. Therefore, when AGN are
present, it is inevitable that biases are introduced when selecting a sample
using these lines. In this work, we select post-starburst galaxies based on the
H$\delta$ and [OII] lines; we discussed the bias due to not considering the
H$\alpha$ line above in Section \ref{sec:halpha_ew}.

Objects that have intrinsically post-starburst stellar populations, but have
significant [OII] emission due to AGN/LINERs \citep{Yan_etal_2006} will be
removed from the sample by the [OII] equivalent width cut. The degree of this
bias is harder to quantify, since it can be difficult to distinguish galaxies
that have post-starburst population with AGN and those with ongoing star
formation, at least at the resolution and signal-to-noise ratio of SDSS
spectra.

We note that the sample of post-starburst galaxies presented in this work is
defined in a practical way, based on the equivalent widths of two lines.
Therefore, it does not correspond exactly to the galaxy population whose
stellar populations are intrinsically post-starburst, regardless of the
presence of an AGN. In the luminosity function calculation in the next section,
we attempt to carefully correct for these biases and check for self-consistency
when possible.

Another source of bias is the incompleteness in the SDSS spectroscopic sample
itself. About 5\% of objects targeted for spectroscopy do not have usable
spectroscopic data in the end. This effect is largely dominated by fiber
collision, with a small fraction due to low signal-to-noise ratio that prevents
redshift measurements. This incompleteness is a weak function of redshift and
magnitude, since faint objects at high redshift tend to have lower
signal-to-noise ratio. All luminosity functions derived in this paper are
divided by $0.95$ to correct for this incompleteness.

\section{Luminosity Function}
\label{sec:luminosity_function}

\subsection{Synthetic Spectrophotometry}
\label{sec:synthetic_spectrophotometry}

In order to calculate the luminosity function of post-starburst galaxies, one
needs the selection function of the sample, i.e., the probability of selecting
a given galaxy as a function of redshift, given its luminosity and spectral
shape. Given that we already have well-calibrated spectra for every object in
our sample, we can synthesize the photometry of an object at any redshift by
shifting the spectrum and scaling the brightness to that redshift,
reconstructing the magnitudes and tracing through the selection criteria. This
process automatically incorporates the k-correction and the dimming of objects
with redshift (the inverse square law). We also quantified and corrected for
the incompleteness due to changes in signal-to-noise ratio as a function of
assumed redshift. The details of these processes are as follows.

We calculate the synthesized magnitude in the $g$, $r$ and $i$ bands that a
given galaxy would have, on a grid of redshift ranging from $z_\mathrm{min}$ to
$z_\mathrm{max}$ (determined as the redshift range where the 3600-4400\AA\,
region would be inside the spectral coverage, as used in the post-starburst
selection method discussed earlier), with a grid spacing of $\Delta z = 0.001$.
For each value on this redshift grid, the spectrum of the object is shifted,
then integrated over each required band to find the average flux density over
the band.

For the vast majority of assumed redshifts, the $g$, $r$ and $i$ filters fall
within the observed spectral coverage. If not, we extrapolate the spectrum
using the best-fit model from the spectroscopic pipeline
\citep{Bolton_etal_2012}. In rare cases when even the model does not have the
required spectral coverage, the magnitude is calculated from linear
extrapolation from nearby redshift values.

This average flux density is then scaled to take into account the inverse
square law by the multiplicative factor
$\left(\frac{1+z_\mathrm{obs}}{1+z_\mathrm{ass}}\right)
\left(\frac{D_L(z_\mathrm{obs})}{D_L(z_\mathrm{ass})}\right)^2$, where
$z_\mathrm{obs}$ is the observed redshift of the object, $z_\mathrm{ass}$ is
the assumed redshift at this particular gridpoint and $D_L(z)$ is the
cosmological luminosity distance evaluated at these redshifts. The relationship
between $D_L(z)$ and $z$ is from the standard $\Lambda$CDM cosmology with
$\Omega_M = 0.3$, $\Omega_\Lambda = 0.7$ and $H_0 = 70
\,\mathrm{km}\mathrm{s}^{-1}\mathrm{Mpc}^{-1}$. This scaled average flux
density is then turned into AB magnitude, with the
\citet{Schlegel_Finkbeiner_Davis_1998} galactic extinction of the object in the
same band corrected for. This extinction correction assumes the standard
total-to-selective extinction ratio $R_V=3.1$. We call this class of magnitude
derived from the spectrum the ``Spectro'' magnitude, as we described in Section
\ref{sec:sdss_magnitude}.

Next, these Spectro magnitudes in the $g$, $r$ and $i$ bands are converted into
various other kinds of magnitudes that are used in the spectroscopic selection
algorithms (Section \ref{sec:target_selection}), namely the Model, cModel,
Petrosian and Fiber magnitudes (Section \ref{sec:sdss_magnitude}). This
conversion is done for each kind of magnitude by applying the difference
between the Spectro magnitude and that kind of magnitude at the observed
redshift. This single-valued conversion is justified because at the redshifts
relevant for the CMASS Sample, the angular diameter distance is a slowly
changing function of redshift. The appearances of the images observed, except
for brightness, are relatively constant, therefore the difference between
families of magnitudes that arise from different ways to process this image
should not change very much with redshift. Once this process is done, we have
determined various magnitudes in the $g$, $r$ and $i$ bands as a function of
assumed redshift. At the lower redshift range relevant to SDSS I/II sample, the
justification that the angular-diameter distance is a slowly changing function
of redshift is no longer true, and the use of the single-valued correction
might be less justified. However, we still expect that it is largely valid
given the small redshift range ($\Delta z\sim0.2$) involved.

\subsection{Magnitude and Color in Fiducial Bands}
\label{sec:fid_band}

In order to compare intrinsic luminosities of objects over a range of redshift,
the magnitudes in the observed bands are not sufficient, since they sample
different parts of the spectrum at different redshifts, and thus need to be
k-corrected. For this purpose, we calculate the absolute AB magnitude from the
spectra in two fiducial top-hat bands fixed in rest-frame wavelength from the
spectra for all our objects from the SDSS I/II and BOSS samples. The first is
the fiducial band we have introduced before; it lies between 4150\AA\, and
4300\AA, sampling the continuum between the H$\gamma$ and H$\delta$ lines. The
other band is between 4950\AA\, and 5100\AA. As introduced earlier, we will
refer to these bands as the [4200] and [5000] bands respectively. These two
wavelength ranges roughly represent the wavelength coverage for the $r$ and $i$
bands at redshift $z \sim 0.5$, and lie in continuum regions without strong
absorption lines. The magnitude in the [4200] band is available for every
object in our sample, because our selection process guarantees that this
rest-frame wavelength is covered in the spectrum. The same is not true for the
[5000] band, since it is shifted outside the spectral coverage for redshift
greater than around unity. However, galaxies in subsamples with homogeneous
selection that will be used to calculate the luminosity function (the SDSS Main
Galaxy Sample and the BOSS CMASS Sample) do not extend to such high redshift,
so this is not an issue.

For each band, the averaged flux density $\bar{f}_\lambda$ in the band is
calculated from the spectrum, then scaled to 10 parsec by the factor
$(1+z)\left(\frac{D_L(z)}{10\mathrm{pc}}\right)^2$ for the calculation of
absolute magnitude in the AB system. This Spectro magnitude is then converted
to cModel magnitude using the observed difference between cModel and Spectro
magnitude from the SDSS bands, and interpolated to the wavelength where the
fiducial band under consideration lies at the observed redshift. Galactic
extinction is corrected in the same manner, using the extinction in SDSS bands
interpolated to the appropriate wavelength where the fiducial band lies.

\begin{figure}
\begin{center}
\includegraphics[trim=20 10 10 20,clip,width=0.475\textwidth]{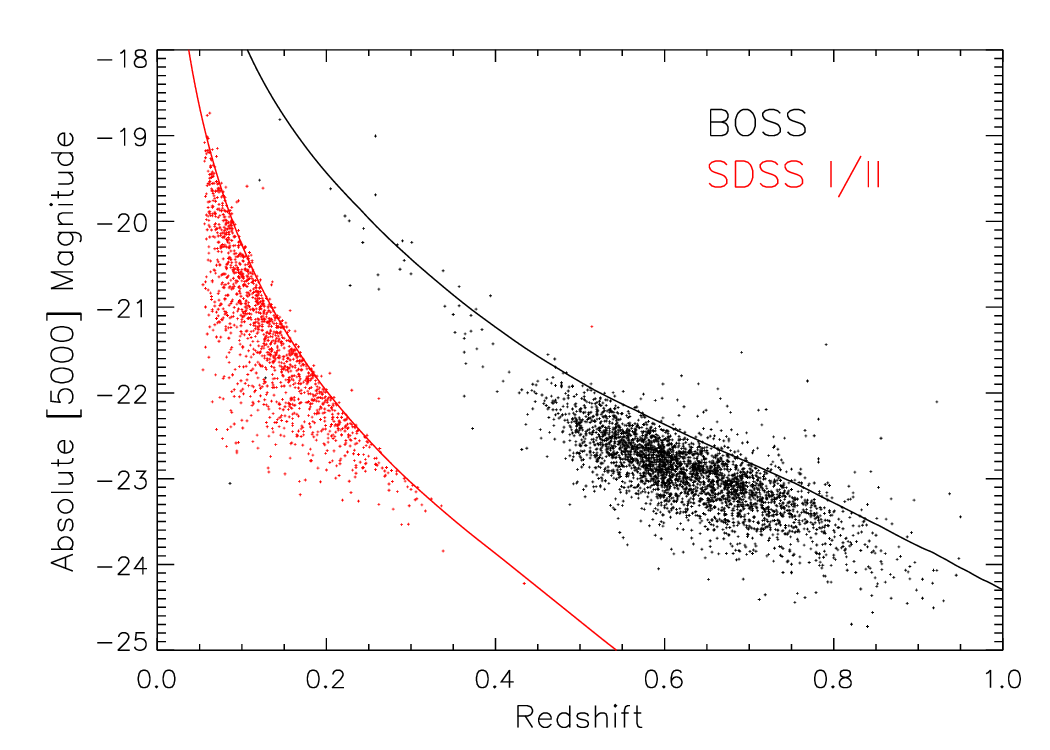}

\caption{Absolute magnitude in the fiducial band [5000] against redshift for
our post-starburst galaxies from both the SDSS Main Galaxy Sample (red) and the
BOSS CMASS Sample (black). The solid lines show the corresponding selection
limits of each sample for the mean SED ($r_\mathrm{petro} < 17.77$ for SDSS and
$i_\mathrm{cmodel} < 19.9$ for BOSS). These lines are calculated from the
coadded spectrum of the sample, such that they show the absolute magnitude in
the [5000] band when the spectrum is scaled to match the selection limits of
the respective surveys at each value of redshift. Note that these selection
lines are not hard cuts, and the fact that some objects lie above the lines is
due to the variation in the SEDs from the coadded spectrum.}

\label{fig:absfidmag_redshift}
\end{center}
\end{figure}

Figure \ref{fig:absfidmag_redshift} shows the scatter plot between this
absolute magnitude in the [5000] band and redshift of the objects, together
with the respective approximate selection limit, in both the SDSS I/II and BOSS
samples. The selection limits are calculated from the coadded spectrum of the
sample, showing the absolute magnitude corresponding to when the spectrum is
scaled to match the selection limits of the respective surveys. Even before
calculating the luminosity function properly, this plot gives a number of
qualitative insights into the luminosity function of these two subsamples of
post-starburst galaxies.

First of all, there is a dramatic difference between the luminosity ranges over
which the post-starburst galaxies lie at low and high redshift, as seen by
comparing the SDSS Main Galaxy Sample and the BOSS CMASS Sample. The
low-redshift sample ranges from $-19$ to $-23$ magnitude, while the
high-redshift sample ranges from $-22$ to $-24$ magnitude and even brighter. It
is noteworthy that there are very few low-redshift post-starburst galaxies
brighter than $-23$ magnitude, even though if luminous galaxies existed at low
redshift they would certainly have been observed and selected into the sample,
although this could be due to small volume sampled. From this, we expect that
the number density of luminous post-starburst galaxies at high redshift is
significantly higher than that at low redshift; this is a downsizing trend.

This trend can also be seen within the low and high redshift samples
separately. For example, at redshift $z\sim0.1$ there are fewer galaxies at
$-23$ magnitude than at redshift $z\sim0.2$, and also at redshift $z\sim0.55$
there are fewer galaxies at $-23.5$ magnitude than at $z\sim0.75$. Therefore we
expect just from this plot (Figure \ref{fig:absfidmag_redshift}) that the
luminosity functions of post-starburst galaxies as a function of redshift
should show a downsizing trend. In other words, luminous post-starburst
galaxies are more abundant at higher redshift. Quantifying this requires taking
into account the selection effects and the fact that the effective volume of
the survey is larger at higher redshift, as we will do in what follows.

It is also noteworthy that the post-starburst galaxies selected from the BOSS
CMASS Sample have systematically higher redshift than the overall CMASS Galaxy
Sample; the median redshift of CMASS post-starburst galaxies is $z\sim0.63$,
while that of the whole CMASS Sample is $z\sim0.57$. This is likely due to
systematic differences in the SEDs of post-starburst galaxies and the generic
massive early-type galaxies which are the main target of CMASS color selection.

\subsection{SDSS Main Galaxy Sample}
\label{sec:sdss_lf}

The SDSS Main Galaxy Sample is simply magnitude-limited at $r_\mathrm{petro} <
17.77$. For each of our post-starburst galaxies from the SDSS Main Galaxy
Sample, we use the Petrosian magnitude as a function of assumed redshift
calculated from the spectrum (see Section
\ref{sec:synthetic_spectrophotometry}) to determine the redshift range over
which this galaxy would be selected into the SDSS spectroscopic sample. Because
our spectra extend to 3800 \AA, we can only select post-starburst galaxies at
$z>0.02$. 

We also calculate the effects of finite signal-to-noise ratio on our selection
function for each galaxy using Monte-Carlo simulations. This completeness
fraction, $\curlyc(z)$, is defined as the fraction of realizations, out of 1000
total realizations at each assumed redshift, that a specific spectrum is
selected as a post-starburst galaxy by our selection criteria (A/Total$>$0.25,
H$\delta$EW $>$ 4.0\AA\, and [OII]$\lambda$3727EW $>$ -2.5\AA) after adding
random noise at the appropriate level. In this calculation, we use the best-fit
template between rest-frame 3600-4400\AA\, range, scaled for cosmological
dimming, as the ``noiseless model'', while the noise is taken from the sky
background at the wavelength range corresponding to rest-frame 3600-4400\AA\,
at that redshift. This noise model is appropriate because at these magnitudes,
the sky noise completely dominates over the photon noise from the object
itself. For majority of objects, the completeness fraction $\curlyc(z)$ remains
near unity over the relevant redshift ranges. The cosmological volume
$V_\mathrm{max}$, within which the object can be observed, is then calculated
as

\begin{equation}
V_\mathrm{max} = \frac{\Omega}{4\pi}\int_{z_\mathrm{min}}^{z_\mathrm{max}} \curlyc(z)\,dV,
\label{eqn:vmax_eqn}
\end{equation}

\noindent where $\Omega$ corresponds to the solid angle of the SDSS survey up
to DR7, 8032 deg$^2$. In cases that we calculate the luminosity function in a
redshift bin, this volume $V_\mathrm{max}$ is further restricted to be only the
volume that overlaps with that redshift bin.

The luminosity function of post-starburst galaxies from the SDSS Main Galaxy
Sample is calculated using the standard $1/V_\mathrm{max}$ method. The sample
is split into two redshift bins at redshift cut $z=0.12$, resulting in roughly
the same number of objects in each bin. We impose an upper limit on redshift
for the $z>0.12$ bin at $z=0.35$, which essentially contains all the
post-starburst galaxies selected from the Main Galaxy Sample. The luminosity
functions are then divided by a factor of two at magnitudes fainter than
$M\sim-22$ to statistically take into account the contamination from objects
with strong H$\alpha$ emission (see Section \ref{sec:halpha_ew} and Figure
\ref{fig:halpha_luminosity}). It is worth noting that this correction may
overcorrect for the contamination at faint magnitudes, since some objects might
have intrinsically post-starburst stellar populations, but with AGN causing
their removal from the sample. The resulting luminosity functions for galaxies
in this sample for both redshift bins (along with that of the CMASS Sample, see
next section) are shown in Figure \ref{fig:lf_redshift}.

\subsection{BOSS CMASS Sample}
\label{sec:boss_lf}

For the BOSS CMASS Sample, the spectroscopic target selection criteria are
designed to select an approximately volume-limited sample of galaxies in the
range $0.4<z<0.7$ to measure Baryon Acoustic Oscillations in their clustering
\citep{Anderson_etal_2012, Reid_etal_2016}. CMASS stands for ``Constant Mass'',
and the following criteria are designed to be crude cuts in both photometric
redshift and stellar mass, isolating objects with mass $>10^{11}M_\odot$. These
selection criteria are

\begin{equation} 17.5 < i_{\mathrm{cmod}} < 19.9 \end{equation}
\begin{equation} r_{\mathrm{mod}} - i_{\mathrm{mod}} < 2.0 \end{equation}
\begin{equation} d_{\perp} = r_{\mathrm{mod}} - i_{\mathrm{mod}} -
             (g_{\mathrm{mod}} - r_{\mathrm{mod}})/8.0 > 0.55 \end{equation}
\begin{equation} i_{\mathrm{fib2}} < 21.5 \end{equation}
\begin{equation} i_{\mathrm{cmod}} < 19.86 + 1.6(d_{\perp} - 0.8), \end{equation}

\noindent with two additional star-galaxy separation cuts

\begin{equation} i_{\mathrm{psf}} - i_{\mathrm{mod}} > 0.2 + 0.2(20.0 - i_{\mathrm{mod}}) \end{equation}
\begin{equation} z_{\mathrm{psf}} - z_{\mathrm{mod}} > 9.125 - 0.46z_{\mathrm{mod}}. \end{equation}

The star-galaxy separation cuts are used to select extended sources by
comparing the PSF magnitudes to the model magnitudes. Due to the slowly varying
nature of the angular diameter distance with redshift for the relevant redshift
range of this sample, we expect that any object that satisfies the star-galaxy
separation at its observed redshift and enters the CMASS Sample in the first
place would also satisfy these conditions over the entire redshift range.
Therefore we do not consider these cuts from our calculation of the selection
function.

However, some post-starburst galaxies may be unresolved, and thus will not be
targeted at all by CMASS. Indeed, $\sim$2\% of our objects are targeted as
unresolved quasar candidates, but this fraction is small enough not to alter
the luminosity function results significantly. We therefore calculate the
luminosity function only from objects selected as CMASS galaxies.

\begin{figure}
\begin{center}
\includegraphics[trim=15 5 10 20,clip,width=0.475\textwidth]{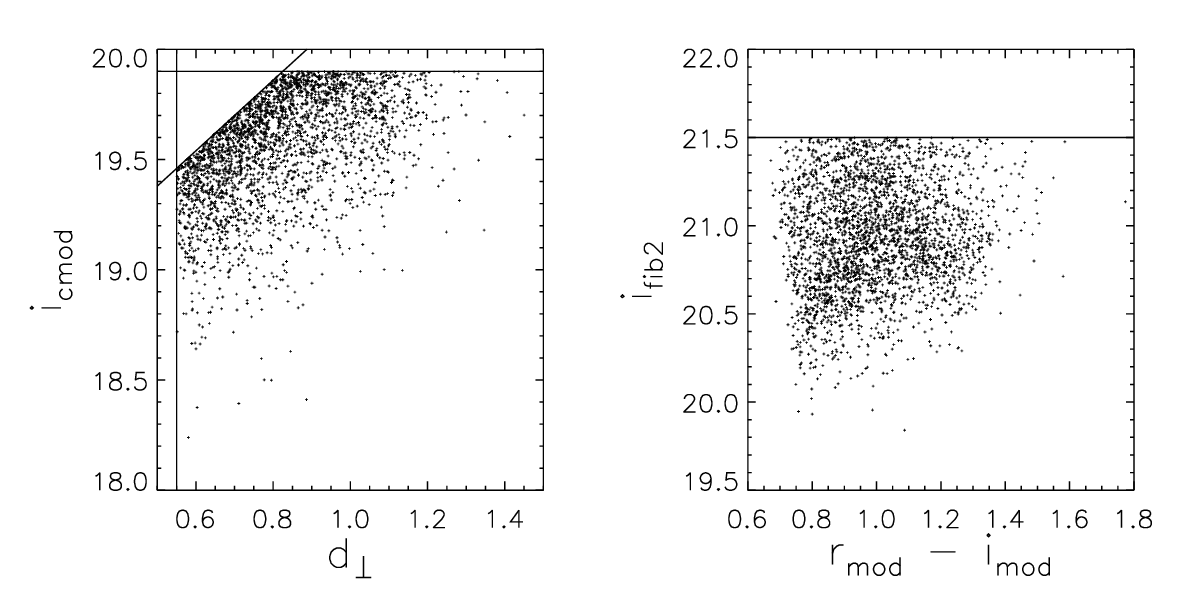}

\caption{Our post-starburst galaxies from the BOSS CMASS Sample shown against
the CMASS selection criteria. The left panel shows the distribution of the
cModel $i$-band magnitude against the color $d_\perp$. The lines shown are the
relevant selection criteria $i_{\mathrm{cmod}} < 19.86 + 1.6(d_{\perp} - 0.8)$,
$i_\mathrm{cmod} < 19.9$ and $d_\perp > 0.55$, with $i_\mathrm{cmod} > 17.5$
off the scale. The right panel shows the Fiber2 $i$-band magnitude against the
$r-i$ color. The selection line shown is $i_\mathrm{fib2}<21.5$, with
$r_\mathrm{mod} - i_\mathrm{mod} < 2$ off the scale. All colors and magnitudes
plotted here are corrected for the \citet{Schlegel_Finkbeiner_Davis_1998}
galactic extinction.}

\label{fig:cmass_selection}
\end{center}
\end{figure}

The distribution of objects in our sample in the various color-magnitude spaces
probed by these criteria are shown in Figure \ref{fig:cmass_selection}. It can
be seen that the three most restrictive conditions for our samples are
$i_\mathrm{cmod} < 19.9$, $d_\perp > 0.55$ and $i_\mathrm{cmod} < 19.86 +
1.6(d_\perp - 0.8)$.

Another potential selection bias that should be noted is the fact that our
selection method is designed to target relatively older post-starburst galaxies
(older than $\sim$50 Myr, the lifetime of a B3 star, which is the least massive
star that produces [OII] emission in the surrounding HII region), leading to
incompleteness in the blue end of the sample (as discussed earlier in Section
\ref{sec:ew_cut}). For the BOSS CMASS subsample, which is designed to target
red, luminous galaxies through this color selection, this effect might add
extra incompleteness to the blue, faint end of the sample.

\begin{figure}
\begin{center}
\includegraphics[trim=30 30 15 30,clip,width=0.475\textwidth]{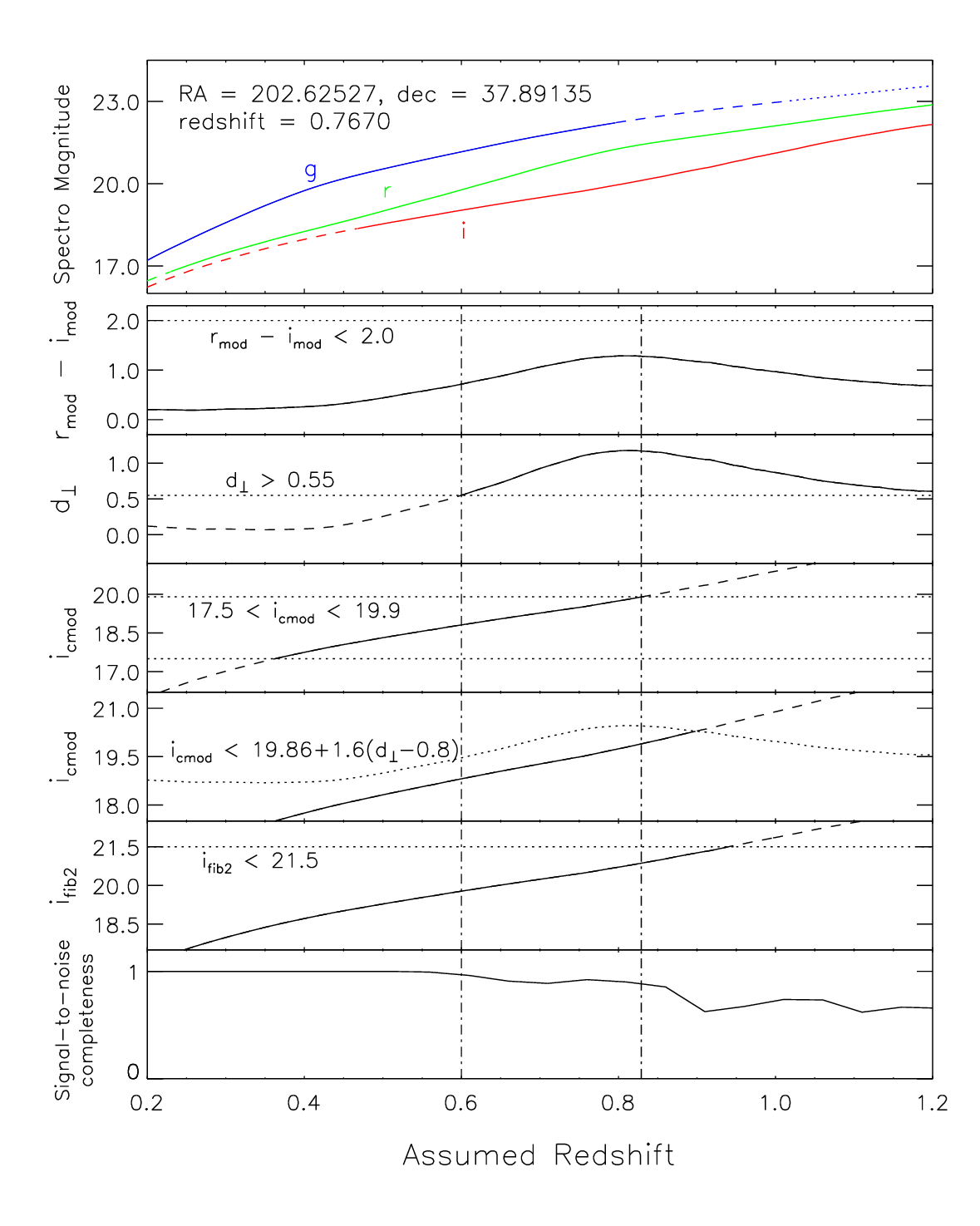}

\caption{Example to demonstrate our method to calculate $V_\mathrm{max}$ for a
specific BOSS CMASS galaxy. The spectrum of this object is shown in Figure
\ref{fig:fitprocedure}. The top panel shows the Spectro magnitude in each of
the $g$, $r$ and $i$ bands that the object would appear to have at any assumed
redshift. The solid lines correspond to the redshift ranges over which the
magnitude can be calculated directly from the spectrum. Dashed lines correspond
to the redshift ranges where the magnitudes are calculated from the best-fit
template because the wavelength coverage of the spectrum does not cover the
band. The dotted lines correspond to the redshift range where neither the
spectrum or the template fit covers the band, and the magnitude is simply
extrapolated. The lower five panels show relevant selection criteria used by
CMASS evaluated in this redshift range, with the thresholdsshown as dotted
lines where relevant. Various kinds of magnitudes are converted from the
Spectro magnitude, with extinction taken into account. The relevant quantities
in each panel are shown by the solid line in the redshift range where the
criteria is met, and dashed line otherwise. The bottom panel shows the
signal-to-noise completeness $\curlyc(z)$, which is derived using Monte-Carlo
simulation, with details described in the text. The horizontal dotted lines
show the selection thresholds. The vertical dash-dotted lines indicate the
redshift range where all five selection criteria are met simultaneously,
corresponding to the redshift range within which this object will still be
selected into the CMASS Sample.}

\label{fig:vmaxcalc}
\end{center}
\end{figure}

We use the prediction of brightness as a function of assumed redshift based on
the observed SED (see Section \ref{sec:synthetic_spectrophotometry}) to find
the redshift range over which all criteria are satisfied, as demonstrated in
Figure \ref{fig:vmaxcalc}. This range corresponds to the redshifts that the
object would still be selected into the spectroscopic sample and subsequently
our post-starburst galaxy sample. We also take into account the effect of
finite signal-to-noise ratio by using Monte-Carlo simulations as described in
Section \ref{sec:sdss_lf}, by adding appropriate random noise based on the sky
background and measuring the A/Total ratio and H$\delta$ and [OII] equivalent
widths for each noise realization, resulting in the completeness fraction
$\curlyc(z)$. The behavior of $\curlyc(z)$ as a function of assumed redshift,
which is more complete at lower redshift and starts to degrade at higher
redshift due to lower signal-to-noise ratio, is typical of the objects in the
sample. The cosmological volume a galaxy can be observed $V_\mathrm{max}$ is
calculated using Equation \ref{eqn:vmax_eqn}, with the solid angle
corresponding to SDSS DR9 area of 3275 deg$^2$. For a redshift bin, the volume
$V_\mathrm{max}$ is further restricted to be only the overlapping volume with
that redshift bin.

Even after the volume $V_\mathrm{max}$ is known, simply summing the inverse of
this volume in bins of magnitudes is not sufficient to get an unbiased
luminosity function. This is due to a consequence of the CMASS color selection
that not all SEDs are selected at all luminosities. We now describe the process
we adopt to correct for this effect.

\begin{figure}
\begin{center}
\includegraphics[trim=5 0 10 20,clip,width=0.475\textwidth]{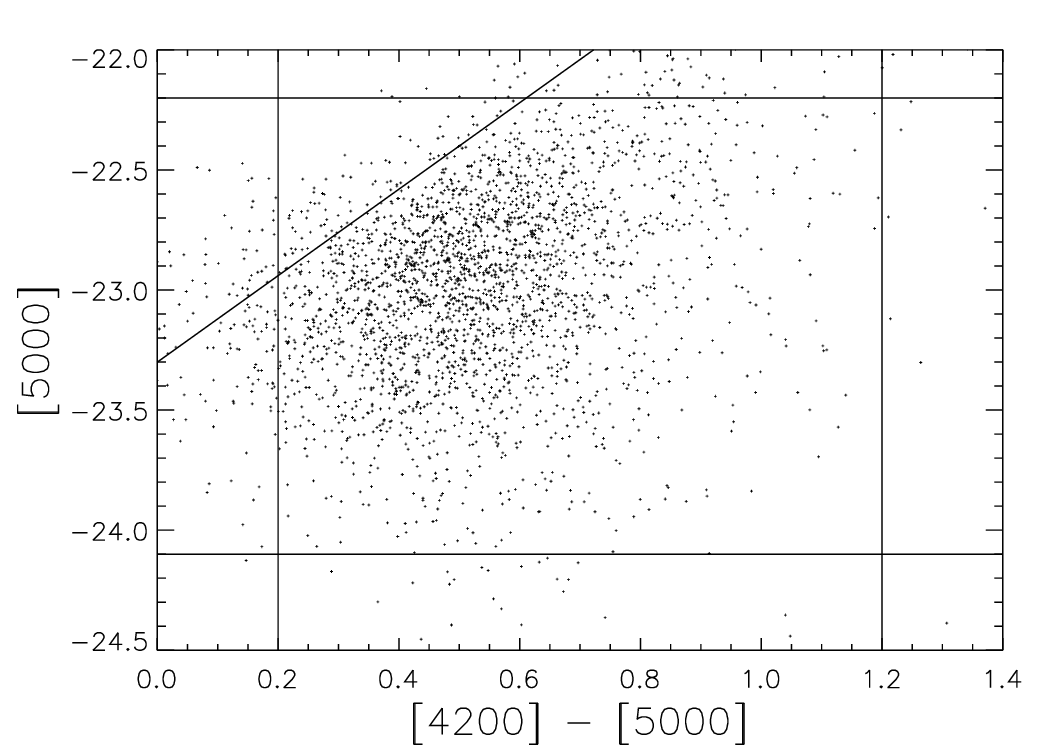}

\caption{Our post-starburst galaxies from the BOSS CMASS Sample at all
redshifts, shown in rest-frame absolute magnitudes and colors calculated from
the spectra. The solid lines shown are the selection cuts in this parameter
space corresponding roughly to the CMASS selection in term of $i$ and $d_\perp$
at redshift $z\sim0.5$. At magnitude fainter than $M\sim-23$, the
color-dependent magnitude cut introduces an additional selection bias in the
luminosity function calculation.}

\label{fig:intrinsic_selection}
\end{center}
\end{figure}

\begin{figure}
\begin{center}
\includegraphics[trim=5 15 30 20,clip,width=0.475\textwidth]{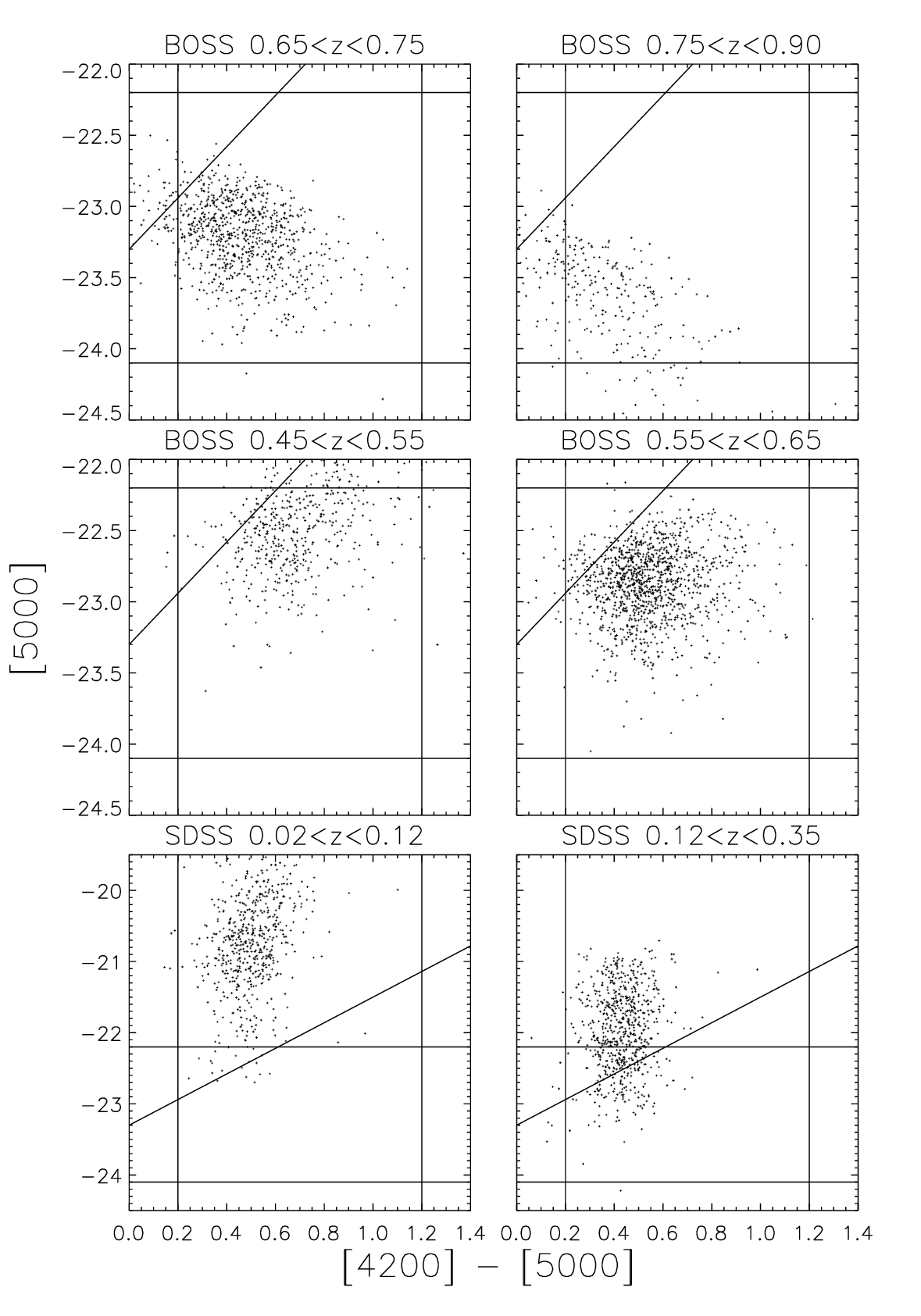}

\caption{Distribution of our post-starburst galaxies in rest-frame magnitudes
and colors similar to Figure \ref{fig:intrinsic_selection}, but separated into
redshift bins. The top four panels show the BOSS CMASS Sample, along with the
selection cuts in these quantities. Only objects that pass these selection cuts
are used to calculate the luminosity functions. The lower two panels show the
equivalent plot for the SDSS I/II Main Galaxy Sample. The selection cuts used
in BOSS sample are shown only for comparison for these two panels, but are not
used since this sample is not subject to the CMASS selection.}

\label{fig:intrinsic_selection_lum}
\end{center}
\end{figure}

Figure \ref{fig:cmass_selection} shows the selection of the CMASS galaxies in
observed quantities, but in order to obtain a physically meaningful measure of
the luminosity function, we need to work in terms of the intrinsic, rest-frame
properties of the galaxies. We measure these intrinsic quantities by using the
two fiducial bands at 4200\AA \, and 5000\AA \, that we introduced earlier.
Note again that these two bands lie at the approximate wavelengths of the $r$
and $i$ bands at redshift $z=0.5$, and the quantity $d_\perp$ is dominated by
$r$ and $i$. Therefore we expect that the selection criteria in terms of
$i_\mathrm{cmod}$ and $d_\perp$ should translate roughly to cuts in the
$[5000]$ and $[4200]-[5000]$ plane. Figure \ref{fig:intrinsic_selection} shows
this plane; galaxies are indeed roughly bounded by the cuts shown. We see that
the color distribution of galaxies in this space becomes increasingly
restricted fainter than $M=-23.0$, which would introduce biases in the
luminosity function if not corrected for. In this parameter space, we define
the selection lines to be $-24.1 < [5000] < -22.2$, $0.2 < [4200] - [5000] <
1.2$ and $[5000] < -23.3 + 1.8 ([4200]-[5000])$. These selection lines are
drawn arbitrarily to have a similar form to the CMASS selection in observed
quantities, with values such that they roughly represent the distribution in
Figure \ref{fig:intrinsic_selection}. In order to have a uniformly selected
sample, we do not include any galaxy that falls outside this range in our
luminosity function analysis. Figure \ref{fig:intrinsic_selection_lum} shows
the distribution in this parameter space but separated into different redshift
bins. For the BOSS CMASS Sample, populations at different redshifts fall into
different region of this parameter space, but remain bound by the fiducial
selection cuts.

\begin{table*}
\begin{center}
\begin{tabular}{|c|c|c|c|c|}
\hline
Survey            & Redshift Range & Total Number & After Fiducial  & Median Redshift \\
                  &                &              & Selection Cuts  &                 \\
\hline
SDSS Main Galaxy  & All Redshifts  &     1462     &      1462       &      0.124      \\
SDSS Main Galaxy  & $0.02<z<0.12$  &      696     &       696       &      0.091      \\
SDSS Main Galaxy  & $0.12<z<0.35$  &      766     &       766       &      0.160      \\
\hline                                         
BOSS CMASS Galaxy & All Redshifts  &     3475     &      2867       &      0.624      \\
BOSS CMASS Galaxy & $0.45<z<0.55$  &      649     &       437       &      0.521      \\
BOSS CMASS Galaxy & $0.55<z<0.65$  &     1462     &      1331       &      0.599      \\
BOSS CMASS Galaxy & $0.65<z<0.75$  &     1039     &       900       &      0.689      \\
BOSS CMASS Galaxy & $0.75<z<0.90$  &      325     &       199       &      0.786      \\
\hline
\end{tabular}

\caption{Details of redshift bins in the luminosity function calculation of the
BOSS CMASS Sample. The first and second columns are the survey and redshift
range respectively. The third column is the total number of objects selected
into each respective sample in these redshift ranges. The fourth column is the
number of objects actually used in the luminosity function calculation. For the
SDSS I/II sample, these two columns are the same. For the BOSS CMASS Sample,
there is an additional requirement for objects to pass the fiducial selection
cuts defined in term of intrinsic properties described in the text. The last
column shows the median redshift of all objects in the third column.}

\label{tab:number_lf}
\end{center}
\end{table*}

\begin{figure}
\begin{center}
\includegraphics[trim=10 0 20 10,clip,width=0.475\textwidth]{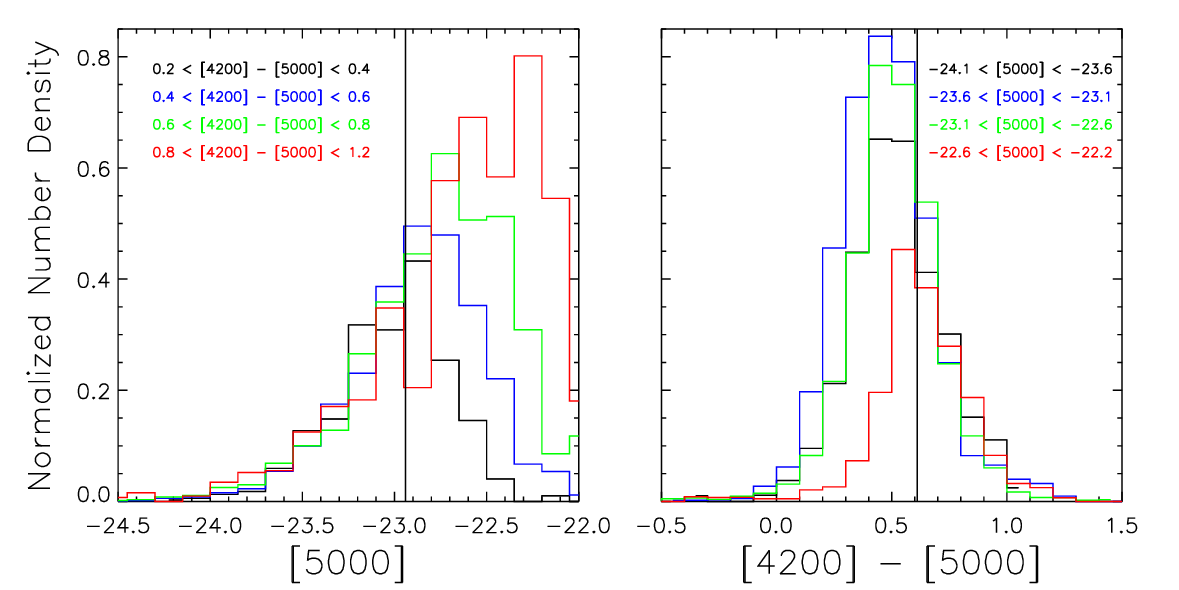}

\caption{The relative distributions of the number density of galaxies
($\Sigma(1/V_\mathrm{max})$) as a function of color and magnitude. Left panel:
the relative distribution as a function of magnitude for different colors,
normalized such that the total numbers of galaxies brighter than
$M_\mathrm{crit}$ (vertical line) is the same for each color bin. Right panel:
distribution as a function of colors for different magnitudes, normalized such
that total numbers of galaxies redder than $C_\mathrm{crit}$ are the same for
each magnitude bin. With this normalization, the curves are approximately
coincident where the distributions are complete, showing that the distributions
of galaxies in color and magnitude are indeed approximately separable and
independent.}

\label{fig:red_complete_vmax}
\end{center}
\end{figure}

This correction works under the assumption that the distribution of number
density of galaxies is a separable function of magnitude and color. Thus
knowing the distribution of galaxies in the red portion of the sample, where
the sample is complete, allows one to extrapolate and determine the number
density in the blue part where the sample is biased against faint objects. The
same argument also applies the other way around, allowing one to use the
luminous portion of the sample, which is complete in color, to determine the
number density at the fainter part where the sample is biased against blue
objects. The validity of this assumption is shown in Figure
\ref{fig:red_complete_vmax}, where the number densities (sum of
$1/V_\mathrm{max}$) for the sample are shown for a number of bins in both
magnitude and color. The distributions in different bins, once normalized, are
the same at the bright and red portions of the sample where the selection is
complete. This property of the distribution allows us to infer the number
density of objects in the region of parameter space where the selection is
incomplete from the region where the selection is complete. There are at least
a few tens of objects in each bin near the cutoff line, therefore small number
statistics should not have a major effect in the conclusion that the number
density is a separable function in color and magnitude.

The correction for this incompleteness is as follows. We define a relevant
color-magnitude parameter space given by $C = [5000] - [4200]$ and $M =
[5000]$. For each gridpoint in this parameter space, we define the distribution
$D(M,C)$ such that $D(M,C)\Delta M \Delta C$ is the comoving number density of
post-starburst galaxies with color between $C$ and $C+\Delta C$ and magnitude
between $M$ and $M+\Delta M$, which is calculated by summing the inverse
$V_\mathrm{max}$ of all galaxies belonging to that gridpoint.

The number density in grid points that are incomplete because of the selection
can then be predicted by extrapolation from the complete part of the parameter
space, under the assumption that the distribution is a separable function in
magnitude and color. The number density in an excluded grid point, written as
$D_\mathrm{incomplete}(M,C)$, can be calculated from the total number density
in three different regions of the parameter space: the part of the parameter
space at the same color but with magnitude bright enough to be complete
[$\int_{-\infty}^{M_\mathrm{crit}} D(M', C) dM'$], the part at the same
magnitude but with color red enough to be complete
[$\int_{C_\mathrm{crit}}^\infty D(M, C') dC'$], and the part that is both red
and bright enough to be complete [$\int_{-\infty}^{M_\mathrm{crit}}
\int_{C_\mathrm{crit}}^\infty D(M', C') dC' dM'$]. Because the underlying
distribution is a separable function in magnitude, the ratio of the first to
the third quantities are the same as the ratio of $D_\mathrm{incomplete}(M,C)$
to the second quantity. This is because the intrinsic shapes of the
distribution as a function of magnitude, up to a normalization factor, do not
change with color. Another equivalent way to think about this is that the ratio
of the second to the third quantities are the same as the ratio of
$D_\mathrm{incomplete}(M,C)$ to the first quantity, because the intrinsic
shapes of the distribution as a function of color also do not change with
magnitude. Either way, this yields the final expression of
$D_\mathrm{incomplete}(M,C)$ as

\begin{equation}
D_\mathrm{incomplete}(M, C) = 
  \frac{\int_{C_\mathrm{crit}}^\infty D(M, C') dC'
        \int_{-\infty}^{M_\mathrm{crit}} D(M', C) dM'}
      {\int_{-\infty}^{M_\mathrm{crit}}
       \int_{C_\mathrm{crit}}^\infty D(M', C') dC' dM'}.
\label{eqn:com_corr}
\end{equation}

Here $M_\mathrm{crit}$ is the magnitude where the diagonal selection line in
Figure \ref{fig:intrinsic_selection} crosses the blue color limit $C=0.2$,
yielding $M_\mathrm{crit}\sim-23$. The value of $C_\mathrm{crit}$ is the color
cutoff above which the selection is complete. This value depends on the value
of magnitude being considered, given by $C_\mathrm{crit}=\mathrm{max}(0.2,
(M+23.3)/1.8)$. For gridpoints that are on the selection cut and are partially
excluded, we calculate the number density distribution by combining the
information from objects in the complete part with the predicted distribution
for the incomplete part, both weighted by the respective areas of their parts
in the cell.

Once the distribution function over the grid has been determined, the
luminosity function can be calculated simply by integrating over the color
\begin{equation} \Phi(M) = \int_{-\infty}^\infty D(M,C')dC' \end{equation}
where $D(M,C)$ is now the distribution in both color and magnitude corrected
for incompleteness from the selection.

\begin{figure}
\begin{center}
\includegraphics[trim=20 0 20 25,clip,width=0.475\textwidth]{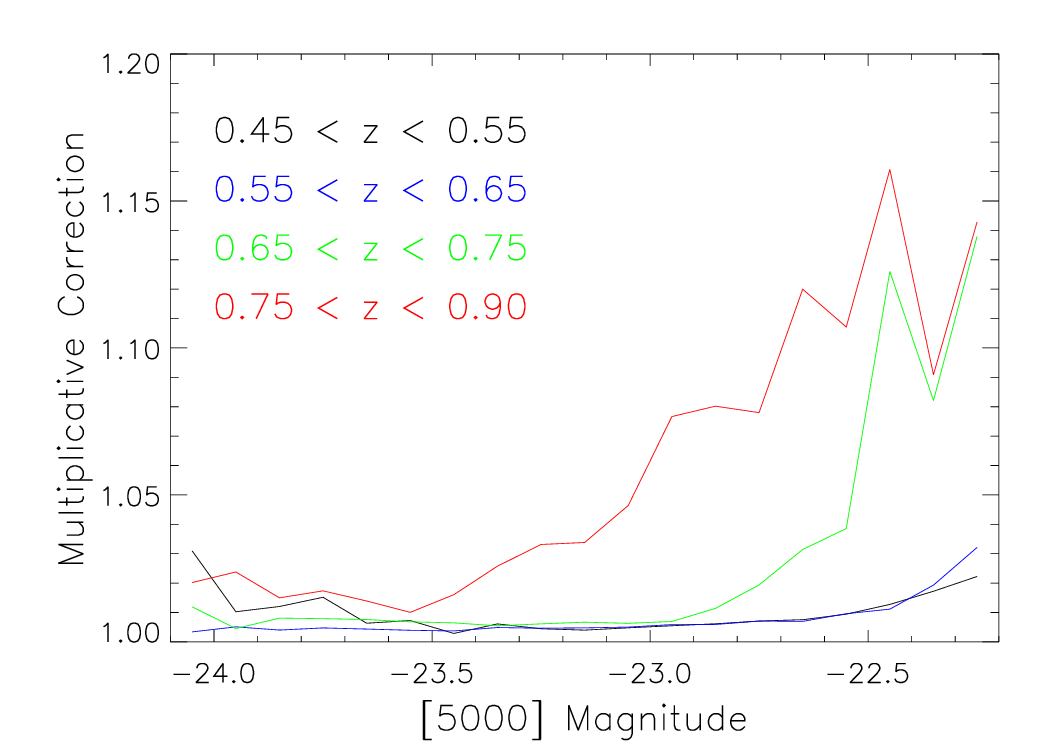}

\caption{The multiplicative correction factor for the BOSS CMASS luminosity
function, to take into account the possible contributions of objects with
signal-to-noise ratio $S/N<2$ which may have been missed by the survey, shown
as a function of redshift bin. This correction factor is calculated from the
fraction of the CMASS galaxies with signal-to-noise ratios less than two in
each respective redshift and rest-frame [5000] magnitude bin.}

\label{fig:sn2_correction}
\end{center}
\end{figure}

Finally, this luminosity function is corrected for another incompleteness due
to possible contribution from objects with low signal-to-noise ratio ($S/N<2$)
that may be missed by our selection method (Figure \ref{fig:signal_noise}) and
thus not in the sample. This correction factor is estimated from the fraction
of the entire BOSS CMASS galaxy sample that have signal-to-noise ratio lower
than 2 in each redshift and rest-frame [5000] magnitude bin used in the
luminosity function calculation. Note that the fraction of CMASS galaxies with
signal-to-noise ratio too low to have redshift measurement is very small. The
resulting multiplicative correction factor for the CMASS luminosity functions
is shown in figure \ref{fig:sn2_correction} for different redshift bins. It
should be noted that the magnitude of this correction is quite small - 20\% in
the most incomplete redshift and magnitude bin. In fact, the most incomplete
bins include no post-starburst galaxies, and thus we do not measure the
luminosity function in those bins anyway. The equivalent correction is not
performed on the SDSS Main Galaxy Sample, since the signal-to-noise ratios of
SDSS galaxies are significantly better than that of CMASS galaxies and we
expect the selection biases due to this effect to be insignificant.

The luminosity function is calculated for our post-starburst galaxies from the
BOSS CMASS Sample in four different redshift bins. We imposed lower and upper
limits to redshift for the lowest- and highest-redshift bins at $z=0.45$ and
$z=0.90$, which contain essentially all post-starburst galaxies selected from
the BOSS CMASS Sample. The number of objects in each redshift bin is shown in
Table \ref{tab:number_lf}. The error bar in the luminosity function is
calculated using Poisson statistics, and propagated through each successive
bias correction factors. However, the error does not take into account possible
systematic errors in the correction for color incompleteness.

\begin{figure}
\begin{center}
\includegraphics[trim=5 0 20 20,clip,width=0.475\textwidth]{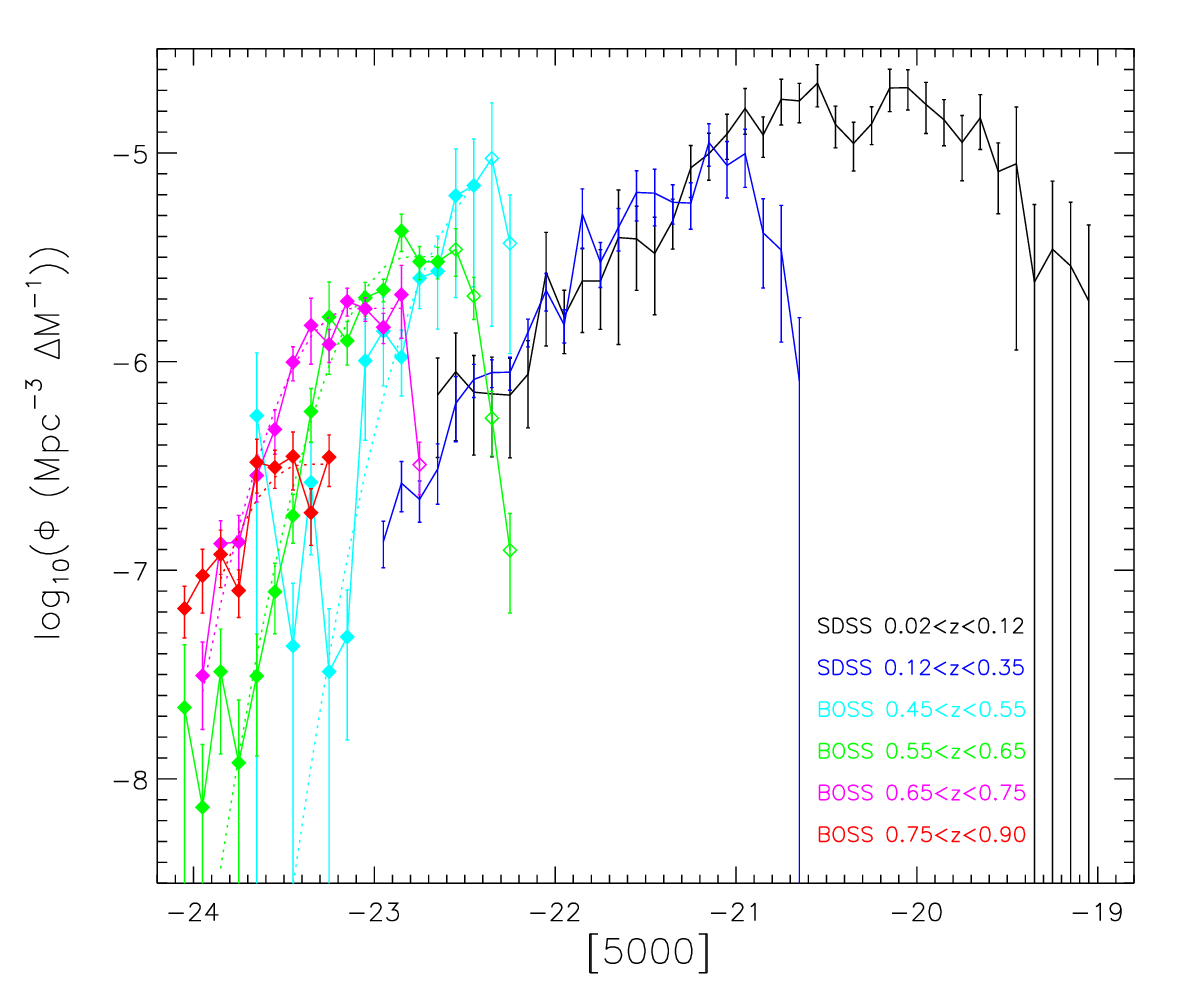}

\caption{Luminosity functions of the post-starburst galaxies from both the SDSS
I/II Main Galaxy Sample and the BOSS CMASS Sample for each of our redshift
bins. For the BOSS CMASS Sample, a luminosity bin indicated by a filled (empty)
symbol has the CMASS color selection correction contributing to less (more)
than 20\% of the value of the luminosity function in that bin. The
half-Gaussian fits (Equation \ref{eqn:lf_fit}) to the luminosity function for
the BOSS CMASS sample are shown as dashed lines. Error bars shown are Poisson.}

\label{fig:lf_redshift}
\end{center}
\end{figure}

Figure \ref{fig:lf_redshift} shows the resulting luminosity functions of BOSS
CMASS post-starburst galaxies for all four redshift bins, as well as those
calculated from the SDSS I/II sample in two redshift bins. The luminosity bins
indicated by filled (empty) symbols in the right panel of the figure have less
(more) than 20\% of their value of luminosity function contributed by this
correction. The SDSS I/II luminosity functions are not shown in empty/filled
symbols to indicate the effect of bias correction, since this correction is
unique to the BOSS CMASS sample due to its color selection.

These luminosity functions are quite different in different redshift bins; we
see strong evidence for redshift evolution. The downsizing trend can be seen in
two ways from this plot. The first is that the magnitude where the luminosity
function peaks becomes progressively more luminous for higher redshift. This
indicates that the typical post-starburst galaxies at higher redshift are more
luminous than at lower redshift. However, selection biases can be important
here, because the peak magnitudes in each redshift bin are very close to the
cutoff magnitudes. We discuss a posteriori tests of the luminosity
function below in Section \ref{sec:self_consistency}.

The second way to see the downsizing trend, which is less affected by the
selection biases, is to compare number densities of post-starburst galaxies at
fixed magnitude for different redshift bins. For example, at $M=-23.5$ the
number densities for different redshift bins gradually decrease from $z\sim0.7$
to $z\sim0.5$, while the two low-redshift bins of SDSS have almost no galaxies
at that magnitude, even though they would have been observed and selected if
they exist. The exception to this trend is the highest-redshift bin
($0.75<z<0.90$) which has the smallest number of galaxies but suffers from the
largest selection biases.

\begin{figure}
\begin{center}
\includegraphics[trim=15 5 20 20,clip,width=0.475\textwidth]{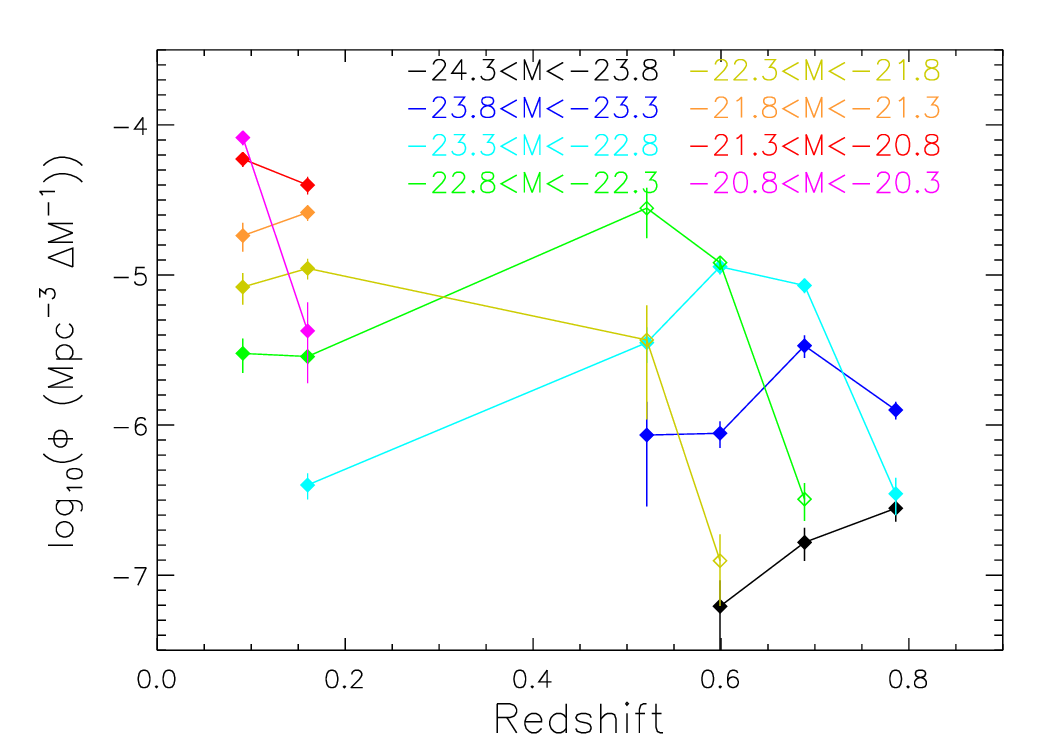}

\caption{Number density as a function of redshift at fixed luminosity. Each
line corresponds to the range of luminosity indicated in the legend, all with
width $\Delta M=0.5$. The error bars are Poisson. This plot contains the same
information as in Figure \ref{fig:lf_redshift} but re-binned in luminosity and
plotted as a function of redshift. As in Figure \ref{fig:lf_redshift}, empty
symbols indicate points affected significantly by the CMASS color selection
correction.}

\label{fig:lf_luminosity}
\end{center}
\end{figure}

This way of viewing the downsizing trend is shown in Figure
\ref{fig:lf_luminosity}. The number densities at fixed luminosity are shown as
a function of redshift. The information used in this plot is the same as that
in Figure \ref{fig:lf_redshift}, using magnitude bins of size $\Delta M=0.5$.
For each luminosity, the number density of post-starburst galaxies increases
strongly with redshift, by more than an order of magnitude from $z=0.1$ to
$z=0.6$ in the case of the sample at $M=-23$, while at fixed redshift, the
fainter objects are more numerous. The exception to these trends is again the
highest-redshift bin and lowest luminosity bin, both of which are likely to be
affected the most by selection effects.

\begin{figure}
\begin{center}
\includegraphics[trim=5 0 15 15,clip,width=0.475\textwidth]{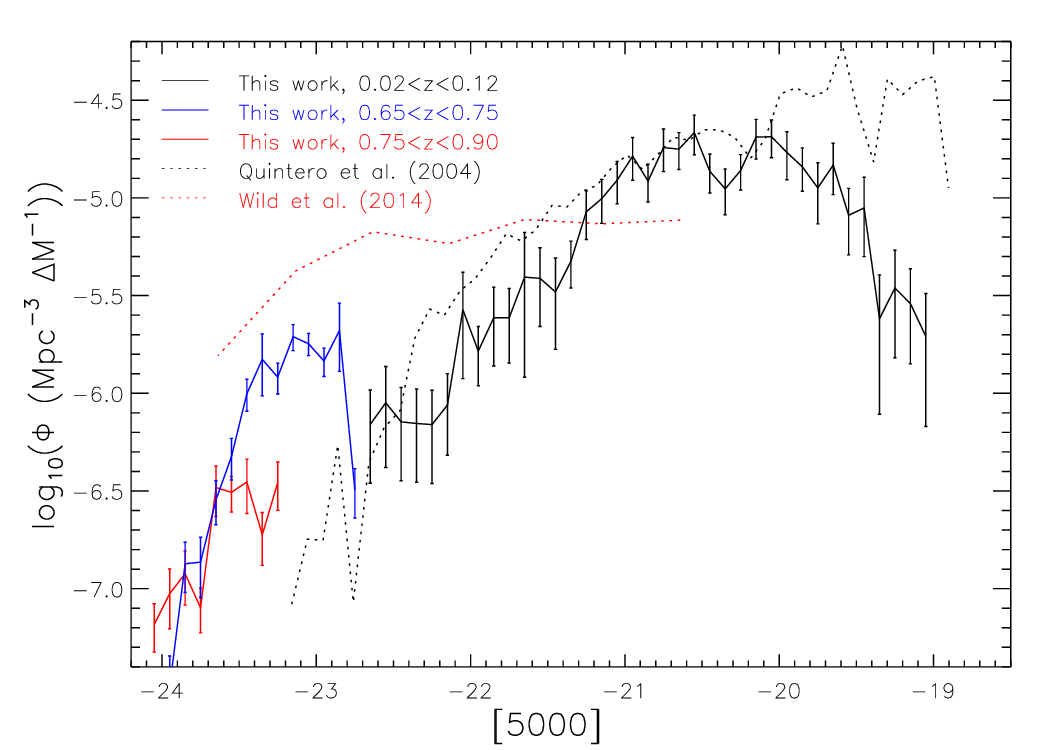}

\caption{Comparison between post-starburst galaxy luminosity functions
calculated in this work to other works in the literature. The solid black, blue
and red lines show the luminosity functions from the SDSS Main Galaxy Sample in
the redshift bin $0.02<z<0.12$ and from the BOSS CMASS Sample in the redshift
bin $0.65<z<0.75$ and $0.75<z<0.90$ respectively. The dotted lines show the
comparable luminosity functions from the literature. The black dotted line is
the luminosity function of post-starburst galaxies in local universe calculated
in the $^{0.1}i$ band, taken from \citet{Quintero_etal_2004}. The red dotted
line is the luminosity function of post-starburst galaxies at redshift $z\sim1$
at rest-frame 1$\mu$m, taken from \citet{Wild_etal_2014}. The unit of $\Phi$
are converted to Mpc$^{-3}$ mag$^{-1}$, with Hubble constant assumed to be
$h=0.7$. The luminosities, measured at different wavelengths in different
papers, are converted to [5000] (rest-frame 5000\AA) using a sensible stellar
population fit that represents a generic post-starburst galaxy spectrum.}

\label{fig:lf_comparison}
\end{center}
\end{figure}

We compare the luminosity functions calculated in this work to that presented
in other works in the literature and show the result in Figure
\ref{fig:lf_comparison}. The lowest redshift bin is compared to the work by
\citet{Quintero_etal_2004}, who calculated luminosity function from
low-redshift SDSS post-starburst galaxies in the $^{0.1}i$-band. The high
redshift comparison is taken from the work by \citet{Wild_etal_2014}, who
calculated the luminosity function of post-starburst galaxies at redshift
$z\sim1$ at the rest-frame 1$\mu$m in the UDS field. We convert all luminosity
functions to a common rest-frame wavelength at 5000\AA\, ([5000] band). We
found that the luminosity functions in our work broadly agree at the
low-redshift bin with what \citet{Quintero_etal_2004} found, but is roughly a
factor of 5-10 lower than that of \citet{Wild_etal_2014} in the overlapping
magnitude range in the highest-redshift bin. Our luminosity function in the
second-highest redshift bin shows more agreement with the result from
\citet{Wild_etal_2014}. The disagreement in the highest-redshift bin may
reflect the selection bias and incompleteness at this redshift. The differences
in wavelength bands, surveys and sample selections likely contribute to this
disagreement as well. It should also be noted that the stellar masses of our
post-starburst galaxies in different redshift bins are likely vastly different.
This is because the color selection of the CMASS sample is designed to isolate
massive galaxies - and is incomplete at redshift $z>0.6$ and
$M_*<10^{11}M_\odot$ \citep{Maraston_etal_2013, Leauthaud_etal_2016} - while
the SDSS Main Galaxy Sample target more intermediate mass galaxies. Our
selection bias correction should correct for part of the CMASS stellar mass
incompleteness, but since we are working in term of absolute magnitude instead
of stellar mass, it is not clear how much has been corrected for.

The luminosity functions confirm that luminous post-starburst galaxies are
indeed a rare class of objects, with a space density
$\Phi_\mathrm{PSG}\sim10^{-(6-7)}\, \mathrm{Mpc}^{-3}\mathrm{mag}^{-1}$ at
$M\sim-23$. For comparison, the space density of early-type galaxies at the
knee of their luminosity function is about $\Phi_\mathrm{EG}\sim10^{-3}\,
\mathrm{Mpc}^{-3}\mathrm{mag}^{-1}$ \citep{Bell_etal_2004, Baldry_etal_2004}.

We find that the luminosity functions of the BOSS CMASS galaxies generally
increase toward fainter magnitudes, as is generally the case for other classes
of galaxies. The general shapes, except for the points dominated by large
selection bias correction, are reasonably fit with Gaussian functions in
magnitude (log-normal functions in luminosity). We thus fit the luminosity
functions with half-Gaussians which are constant at magnitudes fainter than the
peak. This functional form does not decrease at the faint end; note that our
luminosity dynamic range extends only a little below the Gaussian peak. This
function has the same number of free parameters as the Gaussian function, and
is written as

\begin{equation}
\Phi(M) = \Bigg \{
  \begin{array}{ll}
    A \exp\left[-\frac{(M-M_0)^2}{\sigma^2}\right] & , M<M_0\\
    A & , M\geq M_0
  \end{array}
\label{eqn:lf_fit}
\end{equation}

\begin{figure}
\begin{center}
\includegraphics[trim=15 15 25 20,clip,width=0.475\textwidth]{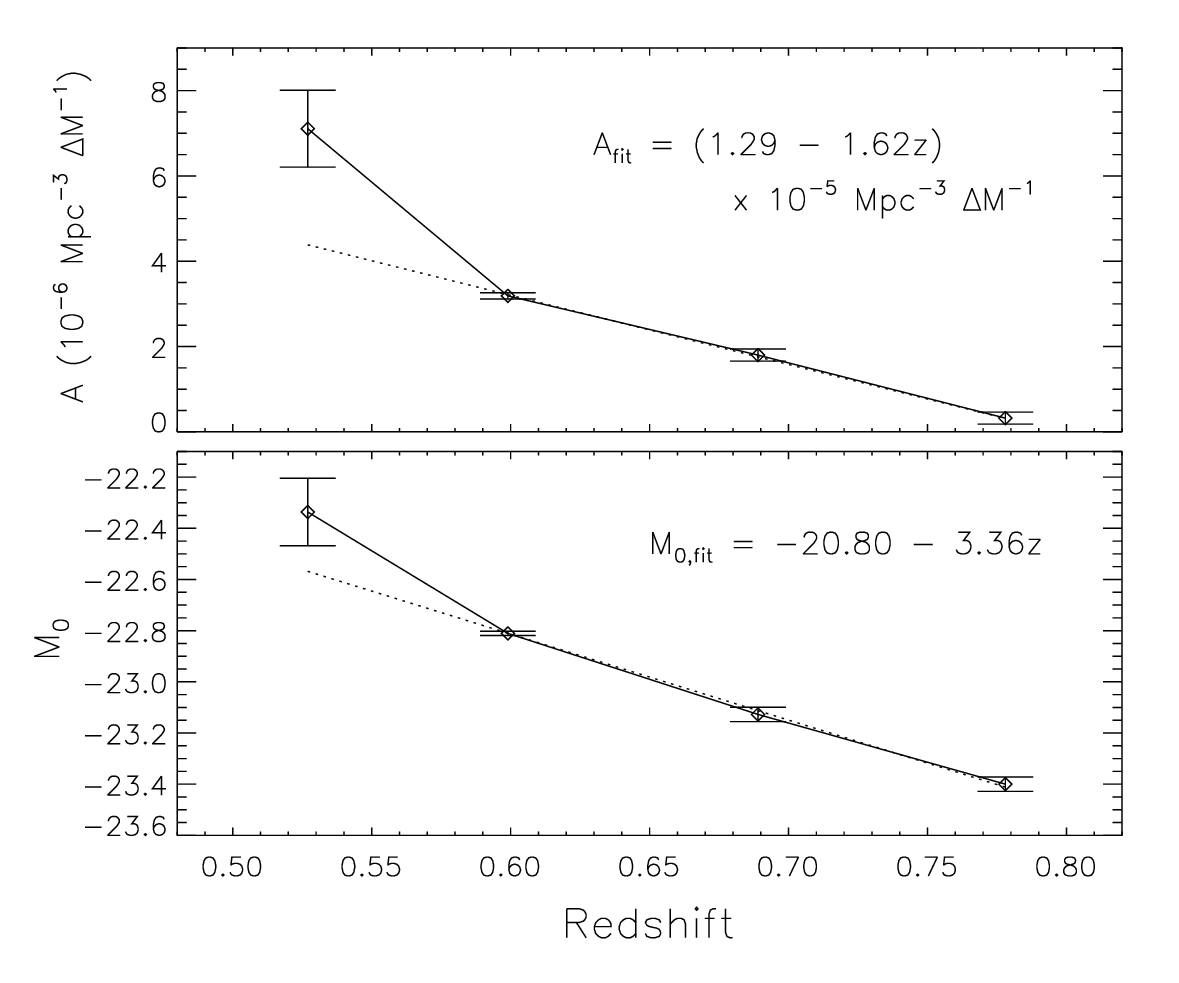}

\caption{Best fit parameters of the luminosity functions of post-starburst
galaxies from our BOSS CMASS Sample for the different redshift bins shown in
Figure \ref{fig:lf_redshift}. The fitting function is half-Gaussian, as
described in Equation \ref{eqn:lf_fit}, where the normalization $A$ and the
peak magnitude $M_0$ are free parameters and the width is fixed as constant at
$\sigma=0.4$. The dotted lines show the linear fit to these parameters. The
redshift on the abscissa is the median redshift of objects in the bin. The
empirical linear fits for these parameters are given in each panel.}

\label{fig:lf_params}
\end{center}
\end{figure}

The two free parameters in this fit are the normalization $A$ and the peak
magnitude $M_0$, while the width of the Gaussian $\sigma$ is fixed to be $0.4$
magnitude at all redshifts; empirically, the width is very close to constant.
With $\sigma$ fixed, the normalization $A$ is proportional to the total number
density in each redshift bin. The fit parameters as functions of median
redshift of the bins, along with a linear fit of these parameters as a function
of redshift, are shown in Figure \ref{fig:lf_params}. This plot shows the trend
that we already described; the overall normalization decreases while the peak
magnitude becomes brighter at higher redshift.

We also attempted to fit the standard Schechter function \citep{Schechter_1976}
to the luminosity function, but we found it to be a poor fit. In the
high-redshift bins, we have insufficient dynamic range to meaningfully
constrain the power-law index, since our magnitude range only samples the
exponential cutoff. Meanwhile, at the low-redshift bins for SDSS I/II, the
luminosity functions do not decrease exponentially at the bright end as the
Schechter function requires.

\subsection{Self-Consistency Test of Luminosity Function from the BOSS CMASS
Sample}
\label{sec:self_consistency}

The validity of the luminosity function is tested in this section with
the method described in, for example, \citet{Sandage_etal_1979} and
\citet{Koranyi_Strauss_1997}. Given the measured luminosity function and the
observed redshift and SED of each galaxy, one can predict what the luminosity
distribution of the galaxies in the sample would be, with color and magnitude
selection taken into account. The predicted luminosity distribution can then be
compared to the observed one. If the sample is uniform and the measurement is
done consistently, the predicted and observed luminosity distributions should
agree well. This method is equivalent to asking the question ``For a given
luminosity function and our selection function, what is the probability
distribution of luminosities for a galaxy of a given redshift and SED shape?''

For each galaxy in the sample, the contribution of this galaxy to the
luminosity distribution bin is given by

\begin{equation}
P_\mathrm{one\ galaxy}(M) \Delta M = 
    \frac{\Phi(M)S(M)\curlyc(M)\Delta M}{\Sigma\Phi(M)S(M)\curlyc(M)\Delta M},
\label{eqn:con_each}
\end{equation}

\noindent where $\Phi(M)$ is the luminosity function. The signal-to-noise
completeness factor $\curlyc(M)$ for each value of $M$ is calculated by a
Monte-Carlo simulation similar to that described in Section \ref{sec:boss_lf},
with the spectrum scaled to that magnitude. The selection function $S(M)$ is
the fraction of the sample that could be observed at magnitude $M$ given the
color cuts. This is determined from the subsample at $M<M_\mathrm{crit}$ where
the sample is complete, which has the same color distribution as the rest of
the sample, as shown in Figure \ref{fig:red_complete_vmax}. At fainter
magnitudes, only a fraction $S(M)$ of this distribution is left due to the
selection cut $M=-23.3+1.8C$. $S(M)$ can be written as

\begin{equation}
S(M) = \frac{\int_{C_\mathrm{cut}(M)}^{\infty} \int_{-\infty}^{M_\mathrm{crit}} D(M',C')dM'dC'}
               {\int_{-\infty}^{\infty} \int_{-\infty}^{M_\mathrm{crit}} D(M',C')dM'dC'}
\end{equation}
where $ C_\mathrm{cut}(M) = (M + 23.3) / 1.8$.

The predicted luminosity distribution is then derived by summing
Equation \ref{eqn:con_each} over the entire sample (or over redshift shells).
The final profile of the predicted distribution integrates to exactly the same
number of objects in the sample, since the normalization for each object is
forced to be unity (as shown in Equation \ref{eqn:con_each}).

\begin{figure}
\begin{center}
\includegraphics[trim=15 0 20 20,clip,width=0.475\textwidth]{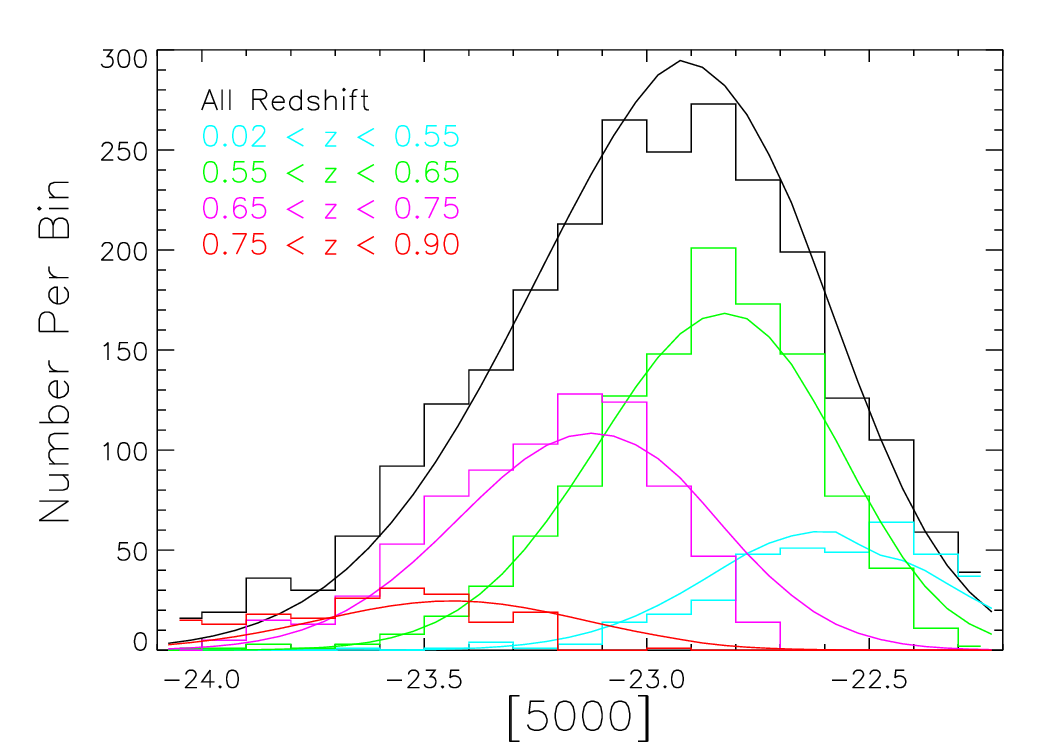}

\caption{Comparison between the predicted and observed luminosity distribution
in the [5000] band of the BOSS CMASS Sample. The histograms show the observed
luminosity distribution (simply numbers of galaxies in each magnitude bin). The
solid lines are the predicted distribution calculated by the method described
in the text. This calculation is performed for both the whole sample (black)
and in redshift bins (colored). The agreement between the two is an a
posteriori confirmation of the validity of our luminosity function
determination.}

\label{fig:lf_selfcon}
\end{center}
\end{figure}

This method in its simplest form assumes an unchanging population, which is in
contrast to the strong redshift evolution we found in Sections
\ref{sec:sdss_lf} and \ref{sec:boss_lf}, and the fast drop in the global star
formation rate of the universe since redshift unity. To address this issue, we
need to take into account the redshift evolution of the luminosity function.
For each galaxy, we evaluate the luminosity function $\Phi(M)$ at its observed
redshift using the fit parameters to the redshift evolution of the luminosity
function shown in Figure \ref{fig:lf_params}. This modified method is performed
for the BOSS CMASS Sample, both for the whole sample and in redshift bins. The
predicted and observed luminosity distributions are shown in Figure
\ref{fig:lf_selfcon}. They agree very well, giving us an a posteriori
confirmation of the robustness of our luminosity function calculation.

It should be noted that, in its simplest form in which both the luminosity
function and luminosity distribution are calculated in a narrow redshift bin,
this method for self-consistency check employs a circular argument. This is
because the underlying population is assumed based on the luminosity function,
then the selection effects are imposed to predict the sample one would observe,
which is exactly the reverse process of calculating the luminosity function in
the first place. However, what breaks the circularity in this case is that now
multiple redshift bins are linked together through the parametrization of the
luminosity function as a function of redshift. The rapid redshift evolution is
not assumed because the luminosity function is calculated individually for each
redshift bin. Therefore the fact that the overall luminosity distribution of
the whole sample, calculated from the evolving luminosity function across 4
redshift bins, agrees with the actual sample is non-trivial.

\subsection{Comparison to Global Star Formation Rate}
\label{sec:toy_model}

We now investigate whether the numbers of post-starburst galaxies in our sample
could explain the decrease in the global star-formation rate since redshift
$z\sim1$. In other words, we test whether all the observed decrease in the
star-formation rate is purely due to quenching of massive galaxies that are in
our sample. In particular, we calculate the mass density in A-stars
($\rho_\mathrm{A}$, in unit of $M_\odot/\mathrm{Mpc}^3$) as a function of
redshift using both approaches.

The first approach is to calculate the A-star mass density from the
post-starburst galaxy luminosity functions, calculated from our sample in
Sections \ref{sec:sdss_lf} and \ref{sec:boss_lf}. This is done by performing
the integral

\begin{equation}
  \rho_\mathrm{A} = \left(\frac{M}{L}\right)_\mathrm{A}
    \left(\frac{A}{\mathrm{Total}}\right) \int\,L\,\Phi(M)\,dM
\end{equation}

This entire calculation is performed in our fiducial top-hat band in the
wavelength range 4950-5100\AA, which we have used in our calculation of the
luminosity function. The integral represents the total stellar luminosity
density in this wavelength band, with luminosity $L$ consistently calculated
from magnitude $M$ in this band. This is then converted to A-star luminosity
density with the global fraction $A/\mathrm{Total}\sim0.26$, which is
calculated from fitting the templates to the coadded spectrum of the entire
sample. Note that this $A/\mathrm{Total}$ value here is calculated in the
[5000] band, and is not the same as the one used in the selection algorithm,
which is in the [4200] band, which explains the low value. Finally it is
converted to A-star mass density with the mass-to-light ratio of A-stars, which
is calculated from the spectrophotometrically-calibrated spectrum of Vega by
\citet{Bohlin_Gilliland_2004}, assuming the distance to Vega of
$7.68\,\mathrm{pc}$.

The second approach is to calculate the A-star mass density expected from the
declining global star-formation rate $\rho_\mathrm{SFR}$. This is done by
integrating the excess star-formation rate backward in time in the time
interval that would result in the stellar population being identified as
post-starburst by our definition at the epoch under consideration.

\begin{equation}
  \rho_\mathrm{A}(t) = \left(\frac{M_\mathrm{A}}{M_*}\right)
  \int\limits_{t-\tau_1}^{t-\tau_2}\,(\rho_\mathrm{SFR}(t') - 
  \rho_\mathrm{SFR}(t-\tau_1))\,dt'
\end{equation}

The global star-formation density $\rho_\mathrm{SFR}$ (in unit of $M_\odot
\mathrm{yr}^{-1} \mathrm{Mpc}^{-3}$) as a function of redshift is calculated
using the fit formula in Equation 15 from \citet{Madau_Dickinson_2014}. The
fraction $M_\mathrm{A}/M_*\sim0.07$ is the ratio of A-star mass to total mass
formed. This is calculated from the Salpeter initial mass function (which is
the same IMF assumed in the formula) assuming the lower and upper limits of
A-star mass to be $1.6-2.9M_\odot$ respectively, while limiting the stellar
masses to be between $0.08-100M_\odot$. The parameters $\tau_1$ and $\tau_2$
are the lower and upper age of the stellar population that would be identified
as post-starburst by our selection algorithm. The lower limit, $\tau_1=50$ Myr,
corresponds to the lifetime of a B3 star, which is the lowest mass star able to
produce an HII region around it, leading to significant [OII] emission. The
upper limit, $\tau_2=800$ Myr, is derived by comparing the single stellar
population model by \citet{Maraston_etal_2005} to our selection, requiring the
A/Total ratio to be larger than $0.25$. This age range, 50-800 Myr, is similar
to the interval between the ages of B-stars and A-stars, as one would
intuitively expect.

One inherent assumption in this estimate is that all galaxies that are still
forming stars at the end of the time window were also forming stars at the
beginning, and therefore are not in the post-starburst population. However, it
can be the case that additional galaxies have quenched during that time period,
joining the post-starburst population, while more star-formation happens in
other galaxies to compensate for the overall star-formation rate. Thus this
estimate is a conservative lower limit, since it only shows the minimum amount
of quenching demanded by the declining global star-formation rate.

\begin{figure}
\begin{center}
\includegraphics[trim=15 15 15 15,clip,width=0.475\textwidth]{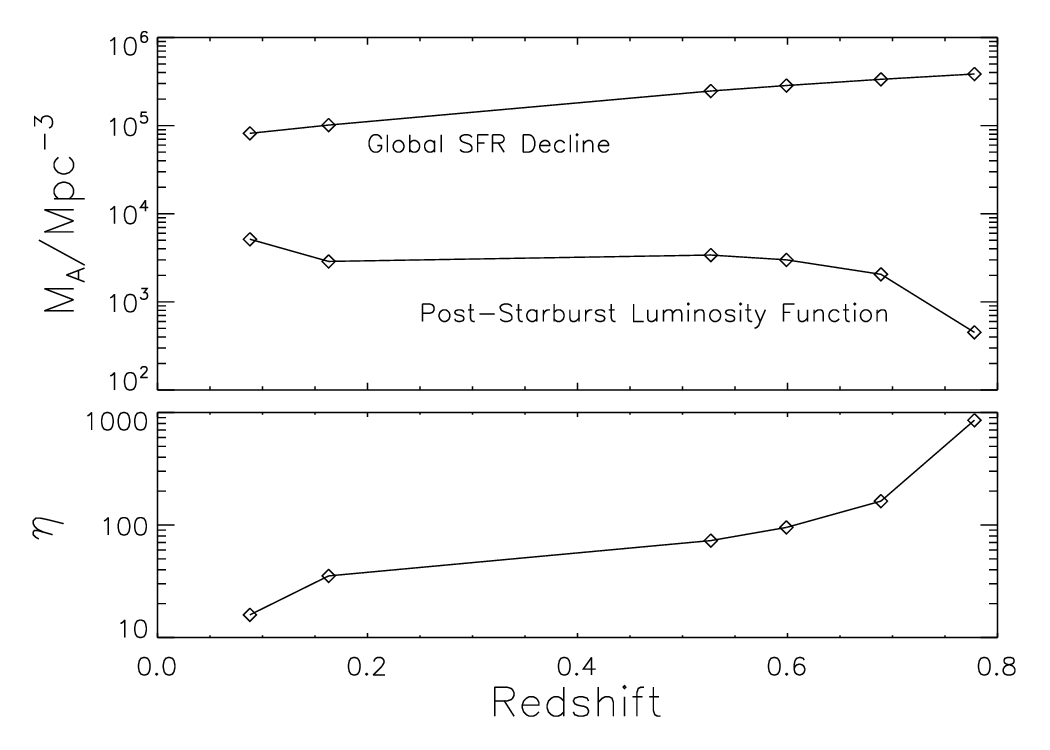}

\caption{Top panel: The mass density in A-stars as a function of redshift (in
units of $M_\odot/\mathrm{Mpc}^3$) calculated in two ways. The bottom line is
calculated from the luminosity functions of post-starburst galaxies in our
sample. The top line is what expected from the declining global star-formation
rate, as quantified by \citet{Madau_Dickinson_2014}. Bottom panel: The ratio of
the densities shown in the top panel. The typical value of around $100$
indicates that the star formation quenching in massive post-starburst galaxies
is a small component of the total decline of star formation in the universe.}

\label{fig:toy_model}
\end{center}
\end{figure}

We performed this exercise of estimating the A-star mass density by these two
different methods in all redshift bins we used to calculate the luminosity
function. This is shown in the upper panel of Figure \ref{fig:toy_model}. We
define the parameter $\eta$, shown in the lower panel of the same figure, as
the ratio between the mass density of A-stars required by the global
star-formation rate to that calculated from luminosity functions. This
parameter, which has typical values of around $100$, suggests that the massive
post-starburst galaxies in our sample account for only a small component of the
decline of star formation in the universe. The vast majority of this decline
must be either in lower luminosity post-starburst galaxies, or in systems in
which the star formation declines only gradually.

Indeed, this calculation contains a number of simplifying assumptions. For
example, the $A/\mathrm{Total}$ ratio, while assumed to be global, can
potentially be a function of luminosity. The IMF is simply assumed to be
Salpeter. The mass-to-light ratio of A-stars are taken from Vega which is an A0
star and therefore is not representative of the whole A-star population. In
addition, the luminosity functions we have calculated may still have residual
selection biases despite all the corrections we have performed, especially in
the highest redshift bin. Each of these assumptions might easily give a factor
of two error to the calculation. However, given that the discrepancy between
the mass density of A-stars calculated from these two different methods is
about two orders of magnitude, our qualitative conclusion is robust to these
uncertain details.

\section{Discussion}
\label{sec:discussion}

\subsection{Sample}

From the publicly available data from the Sloan Digital Sky Survey, we have
identified a large number of post-starburst galaxies based on their optical
spectra. This sample is the largest sample of post-starburst galaxies thus far,
and consists of 2359 galaxies from SDSS I/II and 3964 galaxies from SDSS III
(BOSS) respectively. The highest-redshift objects are at redshift $z \sim 1$
and $z \sim 1.3$ for the two surveys respectively. In fact, for the BOSS
sample, the size of the sample can be increased significantly: the sample used
in this work is selected from the Ninth Data Release (DR9;
\citet{Ahn_etal_2012}), while the BOSS survey now contains twice as many
spectra (DR12; \citet{Alam_etal_2015}). This sample provides a valuable
observational foundation for galaxy evolution studies.

The stellar populations of post-starburst galaxies are modeled with two main
components: young A-stars and an old underlying component. They are selected to
not have any indication of ongoing star formation as measured by
[OII]$\lambda$3727 nebular line emission. So these objects had a large amount
of star formation between 50 million and 0.8 billion years previously that is
almost entirely quenched by the epoch when they are observed.

Their spectra also show a diverse range of features. Weak MgII absorption is
seen in a large number of objects. This feature is related to the intragalactic
gas, and its velocity relative to the systemic redshift indicates inflow or
outflow. A small number of objects show either blue continuum or broad MgII in
emission, both of which are related to AGN activity which is thought to be one
possible feedback mechanism to cause the quenching.

\subsection{Evolution of the Luminosity Function}
\label{sec:lf_evolution}

The luminosity functions calculated from the SDSS I/II and BOSS samples in a
number of redshift bins evolve rapidly over the redshift baseline from
$z\sim0.8$ to $0.1$. The sense of this evolution is consistent with downsizing,
where the typical stellar mass of objects that are currently quenching their
star formation decreases with cosmic time (e.g., \citet{Bundy_etal_2006}).

This dramatic evolution is expected in light of the finding that the global
star formation rate of the universe drops by an order of magnitude since
redshift $z\sim1$ \citep{Madau_Dickinson_2014}. The decrease in global star
formation rate would require a number of star-forming galaxies to be quenched.
These galaxies would subsequently enter the post-starburst phase, in the
redshift range we probe with this survey. Under the assumption that all
galaxies pass through a post-starburst phase after quenching, and that our
sample selection is sufficiently complete, we would expect that the evolution
in the luminosity function of post-starburst galaxies should broadly reflect
the evolution in global star formation rate.

To check this idea, we calculated the mass density in A-stars indicated by the
measured luminosity functions, and compared this to what is expected from the
declining global star-formation rate. We found that the mass density of A-stars
in our post-starburst galaxies is smaller than that expected by a factor of
$\sim100$, suggesting that most star-forming galaxies at high redshift do {\em
not} go through a luminous post-starburst phase, and post-starburst galaxies
are indeed a rare occurrence.

In light of this result, there are a few ways in which the star formation rate
could decrease without entering the post-starburst phase. The first is that
most of the star-formation quenching might happen in low-mass galaxies or
satellites that are not luminous enough to be selected as part of the survey.
The $1/V_\mathrm{max}$ formulation we performed in redshift bins does not
correct for this because the luminosity functions are evolving with redshift,
and thus faint galaxies that are seen locally can not be used to infer the
corresponding population at higher redshift. Indeed, even if there exists a
large population of unseen low-luminosity galaxies at high redshift, it would
not contribute much to the total luminosity and inferred mass. For example, if
the faint end of the luminosity function at $0.65<z<0.75$ mirrored that of the
$0.02<z<0.12$ bin, it would increase the total luminosity in that bin by less
than a factor of 2.

The second way this could happen is for the star formation to be quenched
gradually in the majority of star-forming galaxies, and perhaps staying at a
modest rate thereafter (e.g. \citealt{Cortese_Hughes_2009, Salim_etal_2012,
Fang_etal_2012, Fang_etal_2013, Schawinski_etal_2014}). Such galaxies will
never show a strong A-star component in their spectra, and will not be
identified as a post-starburst galaxy. We believe that this scenario is the
main channel by which the star-formation rate in the universe has decreased,
given that reasonably massive galaxies, like our Milky Way, that retain modest
levels of star formation are common.

Another complication in this aspect is that a fraction of post-starburst
galaxies might not represent a key evolutionary stage of the galaxy, but only a
random, stochastic phase of a small-scale starburst which is not significant
compared to the total stellar mass of the galaxy. Depending on the exact
details of the selection criteria, these objects might pass through the
selection and enter the sample. This issue can be answered, again, by
understanding the mapping between the fundamental properties of the stellar
population to the observable spectral features, which can be done with stellar
population synthesis models.

\subsection{Mechanisms Responsible for Quenching}

Another key aspect is the physical mechanism for quenching the star formation
on a short timescale. There are two main scenarios by which this can happen.
The first involves cluster-related mechanisms such as ram-pressure stripping,
in which a star-forming galaxy falls into a galaxy cluster, and interacts with
the intra-cluster medium. The second is the merger-related scenario, where a
merger between two gas-rich galaxies induces a large starburst which is
subsequently quenched either due to feedback or gas exhaustion. These scenarios
can be distinguished in a number of different ways, some with data available in
this sample, some requiring more complementary information.

One possible way to approach this would be to use stellar population model fits
to the individual galaxy spectra to reconstruct a crude star-formation history
of each galaxy. If the recent star formation history is consistent with a large
starburst that is quenched quickly, then it would be more consistent with the
merger scenario where the large scale starburst is induced all at once,
possibly at the center of the galaxy. In contrast, if the star-formation
history is closer to being constant, before being quenched relatively suddenly,
then it would be closer to the cluster interaction scenario. In this case, the
term post-starburst is certainly misleading, because there is no burst, but it
can have a strong enough A-star feature to be classified under this group if
the star formation is quenched suddenly enough \citep{Quintero_etal_2004}. In
practice, distinguishing these two scenarios in stellar populations is a
challenging task and might only be possible for extreme cases even for a high
signal-to-noise ratio spectra.

Another clue contained within the spectra is the presence of AGN, which
certainly also can play an important role in the evolution of the galaxy, and
might be closely related to how the star formation is quenched. Simulations
\citep{Hopkins_etal_2006, Hopkins_etal_2008, Sijacki_etal_2007} invoke the
quenching from AGN feedback, providing the energy needed to stop overproduction
of stars. However, there is evidence \citep{Wild_etal_2010, Shin_etal_2011,
Yesuf_etal_2014} that AGN activity tends to peak after the starburst end, and
therefore might not be the cause of star formation quenching. Quantifying the
number of post-starburst galaxies with spectroscopic AGN signatures might yield
an understanding of the influence central AGNs have on this population. In our
sample there are $\sim100$ objects that show signs of AGN activity such as blue
continuum or, more importantly, broad line emission. However, this number might
suffer from severe selection bias, since AGN activity can give rise to
[OII]$\lambda$3727 emission, leading to exclusion of objects from our sample.
Potentially, one can also use emission line diagnostics (BPT diagram;
\citet{Baldwin_etal_1981, Veilleux_Osterbrock_1987}) to identify obscured AGN
that do not show broad-line emission. In addition to the optical signature of
the AGN, if complementary data in other wavelength ranges (such as X-ray or
radio) is available then one might also have different ways to detect and
quantify AGN properties and understand this relation (e.g.
\citet{Shin_etal_2011}).

Moreover, this sample can be augmented by a number of external datasets.
Firstly, deep imaging data can be very useful to quantify the environment of
the post-starburst galaxies, i.e., whether they lie in galaxy clusters or the
field. The requirement is that these imaging data have to be deep enough to see
a typical $L_*$ galaxy at redshifts up to unity. The SDSS photometric data are
not deep enough to do this, but various current and upcoming weak-lensing
surveys such as the Canadian-France-Hawaii Telescope Legacy Survey (CFHTLS;
\citet{Heymans_etal_2012}), the Dark Energy Survey (DES; \citet{DES_2005}),
Hyper-Suprime Cam (HSC; \citet{Miyazaki_etal_2012}) and the Large Synoptic
Survey Telescope (LSST; \citet{Ivezic_etal_2008}) would be sufficient for this
task.

High resolution HST imaging data would also be useful in studying the
morphology of these objects at high redshift to probe the effect of close
interactions and mergers, possibly by comparing to a control sample of normal
galaxies with similar mass and redshift. If a significant fraction of
post-starburst galaxies, especially ones in the field, show signs of disturbed
morphology or close, merging neighbors, it would lend support to the merger
scenario. It would also be interesting if one finds field post-starburst
galaxies with smooth morphology, since that would need some other explanation.
There are a number of works studying morphologies at low redshift (e.g.,
\citet{Yang_etal_2008} for post-starburst galaxies, \citet{Cales_etal_2011} for
post-starburst quasars). However, since the post-starburst galaxy population at
high redshift is significantly different from that at low redshift, as seen in
the evolution of the luminosity function, independent studies at high redshift
must be done to fully understand the properties and mechanisms at work in the
high-redshift population.

Indeed, the single most important piece of information that would allow us to
explore different ways that galaxies evolve is the stellar mass. This is
because the observed color and luminosity change in complex ways due to stellar
evolution, while the stellar mass reflects fundamental changes in galaxy
properties like mergers, infall or star formation. In general, the stellar mass
can be determined by spectral fitting up to a factor of a few. However, it
should be noted that the post-starburst stellar population may be dominated by
the relatively less-understood TP-AGB stars, and therefore the error in
determining stellar mass might be larger.

\subsection{Improvement to Selection Method}

In this work, we have developed a robust method to select post-starburst
galaxies from optical spectra. This method relies minimally on prior knowledge
of the redshift of the object, making it suitable to apply to large amounts of
spectroscopic data even in the presence of large redshift errors in the
standard pipeline. This algorithm selects a relatively small number of
candidates from the large dataset for visual inspection to confirm the
correctness of the identification. While the false negative rate is very small
based on a small test sample of known post-starburst galaxies, the false
positive rate to be removed by visual inspection is between 100\% and 300\%. At
this point, the amount of visual inspection required is still manageable, but
more optimized algorithm with lower false positive rate will be critical when
the next generation of surveys generate significantly larger samples. Examples
of such surveys are the fourth phase of the Sloan Digital Sky Survey (SDSS IV;
\citet{Dawson_etal_2015}), The Dark Energy Spectroscopic Instrument (DESI;
\citet{Levi_etal_2013}) and the Prime Focus Spectrograph (PFS;
\citet{Takada_etal_2014}). For SDSS and BOSS data in particular, it was found
in retrospect that the redshift determination by the spectroscopic pipeline is
very robust, with less than 5\% incorrect redshifts even for this rare class of
objects. Therefore, selecting the post-starburst galaxies from SDSS and BOSS
assuming the pipeline redshifts to be reliable would have simplified the
process and reduced the number of false positives significantly.

In light of this, there are some improvements to incorporate the future work
selecting post-starburst galaxies from large spectroscopic surveys. One such
dataset in the near future would be the entire BOSS survey (SDSS DR12). This
phase of SDSS contains twice as many spectra than what we used in this paper,
and is now public \citep{Alam_etal_2015}.

The first improvement might be to do the selection based on the stellar
population properties, rather than empirical cuts on parameters such as A/Total
or equivalent widths. Ideally, this approach is more preferable since it would
lead to selection based on physically meaningful properties of the galaxy
population such as stellar population age and metallicity. For example, one
might define post-starburst galaxies to be populations that are older than 300
Myr but younger than 1Gyr. \citet{Yesuf_etal_2014} fit stellar population
models using multi-wavelength data while taking AGN and dust obscuration into
account.

Selection of these objects from the next generation of multi-band photometric
surveys is another possibility. Doing so requires distinguishing between the
3600\AA\, Balmer break and the 4000\AA\, break at slightly different redshifts.
One can potentially resolve this ambiguity using multi-wavelength information,
for example in NUV and NIR, since the old population would have almost no flux
in NUV but large flux in NIR, and conversely for the young population.
\citet{Wild_etal_2014} studied galaxy SEDs in optical and NIR, and found that
post-starburst galaxies have unique, identifiable colors. However, the presence
of dust and AGN would make this task challenging.

\section{Conclusion}
\label{sec:conclusion}

The post-starburst phase is a rare phase in galaxy evolution, in which a small
number of star-forming galaxies are on their way to becoming passive early-type
galaxies. Therefore, understanding this class of galaxies could yield clues
about how galaxies evolve and the mechanisms that drive evolution, at least for
massive galaxies. Currently, there are many open questions in this topic,
ranging from phenomenological aspects such as the luminosity function and its
evolution with redshift, to the physical mechanisms that cause star formation
to stop and subsequently turn a star-forming galaxy into a post-starburst
galaxy. One key limitation faced by many previous works was the lack of a
large, statistical sample with uniform selection over a broad range of
redshift.

In this paper, we have systematically selected a sample of post-starburst
galaxies from the SDSS DR9 spectroscopic dataset. The method we developed is
based on template fitting, with cuts on equivalent widths applied to various
relevant spectral lines. The selection algorithm is tested and verified against
a smaller sample of post-starburst galaxies known prior to the study. All
objects in the final sample are visually inspected to remove false positives.

We apply this selection scheme to the entire SDSS DR7 spectroscopic database,
and also to the BOSS spectroscopic dataset presented in DR9. This yields 2330
galaxies from the SDSS DR7 dataset, and 3964 galaxies from the BOSS dataset.
This is the largest sample of post-starburst galaxies currently available in
the literature. The full list of these objects is available with this paper.
The redshifts of these galaxies range from local galaxies up to $z\sim1.3$,
with a median redshift of $\sim0.16$ and $\sim0.61$ for the SDSS DR7 and BOSS
subsamples respectively, with little overlap between them. Various interesting
spectral features are seen, such as MgII absorption in a large number of
galaxies, and broad MgII emission and blue continuum in about 100 objects. They
are the so-called post-starburst quasars.

A large fraction of SDSS DR7 sample is selected by the Main Galaxy Sample,
while the dominant spectroscopic selection algorithm for BOSS sample is CMASS.
These two subsamples have uniform magnitude and color selection criteria,
allowing us to calculate luminosity functions in a number of redshift bins.

We quantify and correct for various selection biases in the luminosity function
calculation. The first bias is the different cosmological volume over which
each object can be selected from the spectroscopic sample of the survey. This
is corrected by the standard $1/V_\mathrm{max}$ method, where $V_\mathrm{max}$
is calculated from the individual spectrum and the details of the selection
criteria for each subsample. The second effect is due to the color selection
algorithm used in the CMASS Sample, which is biased against faint blue objects,
even though we expect post-starburst galaxies to also exist in that region of
color space. This is corrected by extrapolation from the complete part of the
color distribution, assuming that the underlying distribution is separable in
color and magnitude, an assumption shown to be consistent with the data.
Moreover, after all the corrections are performed, we compare the observed
distribution of objects in luminosity to the expected one given the luminosity
function. The observed and expected distributions are consistent, suggesting
that our method to calculate the luminosity function is robust.

We see strong redshift evolution of the resulting luminosity function of our
post-starburst galaxy sample. The sense of this evolution is that the number
density at fixed luminosity of post-starburst galaxies is considerably higher
at high redshift. In other words, we see the ``downsizing'' trend in the
luminosity function evolution that has been seen in other galaxy populations.
At fixed redshift, the less luminous objects are more abundant than luminous
objects. This trend, qualitatively, is in line with the expectation from the
fact that the global star formation rate of the universe has dropped by about
an order of magnitude since redshift $z\sim1$.

We performed a quantitative comparison between the mass density of the
post-starburst A-star population from our sample (by integrating over the
luminosity function), to that expected from the declining global star-formation
rate. We found that only a small fraction, approximately $1\%$, of the minimum
star-formation quenching required by the declining global star-formation rate,
can be explained by the amount found in post-starburst galaxies. This makes
post-starburst galaxies a rare phenomenon. It suggests that most star-formation
quenching has to happen in ways that escape our post-starburst galaxy selection
method. This can be either by quenching gradually enough to never show strong
post-starburst spectral signatures, or by quenching in low-mass galaxies or
satellites that are too faint to be selected into the SDSS survey at high
redshift.

\section*{Acknowledgements}

We thank Kevin Bundy, Renyue Cen, Jenny Greene, James Gunn, Robert Lupton, and
Christy Tremonti for useful discussions, insights and comments during the
entire duration of this project. We also thank the anonymous referee for
exceptionally thorough feedback, which has improved the paper significantly.

Funding for SDSS-III has been provided by the Alfred P. Sloan Foundation, the
Participating Institutions, the National Science Foundation, and the U.S.
Department of Energy Office of Science. The SDSS-III web site is
http://www.sdss3.org/.

SDSS-III is managed by the Astrophysical Research Consortium for the
Participating Institutions of the SDSS-III Collaboration including the
University of Arizona, the Brazilian Participation Group, Brookhaven National
Laboratory, Carnegie Mellon University, University of Florida, the French
Participation Group, the German Participation Group, Harvard University, the
Instituto de Astrofisica de Canarias, the Michigan State/Notre Dame/JINA
Participation Group, Johns Hopkins University, Lawrence Berkeley National
Laboratory, Max Planck Institute for Astrophysics, Max Planck Institute for
Extraterrestrial Physics, New Mexico State University, New York University,
Ohio State University, Pennsylvania State University, University of Portsmouth,
Princeton University, the Spanish Participation Group, University of Tokyo,
University of Utah, Vanderbilt University, University of Virginia, University
of Washington, and Yale University. 

This work is partially based on observations obtained with the Apache Point
Observatory 3.5-meter telescope, which is owned and operated by the
Astrophysical Research Consortium.

\bibliography{reference}

\end{document}